\documentclass[sigconf,screen]{acmart}
\newcommand{{\proj}}[0]{VisGuardian}
\usepackage{graphicx}
\usepackage{tikz}
\usepackage{multirow}
\usepackage{multicol}
\usepackage{tabularx}
\usepackage[xcdraw,table]{xcolor}
\usepackage{tabularray}
\usetikzlibrary{trees}
\usepackage{subfig}

\usepackage{booktabs}   
\usepackage{ragged2e}  
\usepackage{array}      

\definecolor{riskLow}{HTML}{FFF0E1}     
\definecolor{riskMedium}{HTML}{FFDAB9}   
\definecolor{riskHigh}{HTML}{FFC07A}    
\definecolor{spatialOffice}{HTML}{E8F6FD}     
\definecolor{spatialBedroom}{HTML}{D1EEF7}    
\definecolor{spatialActivity}{HTML}{BDDEEE}   
\definecolor{spatialPersonal}{HTML}{A4D8E9}   
\definecolor{spatialBathroom}{HTML}{93CBE3}    
\definecolor{spatialLiving}{HTML}{80BBDD}   
\definecolor{typeAppendences}{HTML}{E8F8F5}    
\definecolor{typeMarker}{HTML}{D4EFDF}      
\definecolor{typeClothes}{HTML}{A3E4D7}      
\definecolor{typeSafety}{HTML}{7EC6B7}      
\definecolor{typeDigital}{HTML}{5AB2A5}     
\definecolor{typeUncat}{HTML}{FFFFFF}    

\AtBeginDocument{%
  \providecommand\BibTeX{{%
    \normalfont B\kern-0.5em{\scshape i\kern-0.25em b}\kern-0.8em\TeX}}}

\copyrightyear{2026}
\acmYear{2026}
\setcopyright{cc}
\setcctype{by}
\acmConference[CHI '26]{Proceedings of the 2026 CHI Conference on Human Factors in Computing Systems}{April 13--17, 2026}{Barcelona, Spain}
\acmBooktitle{Proceedings of the 2026 CHI Conference on Human Factors in Computing Systems (CHI '26), April 13--17, 2026, Barcelona, Spain}
\acmPrice{}
\acmDOI{10.1145/3772318.3790288}
\acmISBN{979-8-4007-2278-3/2026/04}

\begin{document}

\title[\proj{}: A Lightweight Group-based Privacy Control\\ Technique For Front Camera Data From AR Glasses in Home Environments]{\proj{}: A Lightweight Group-based Privacy Control Technique For Front Camera Data From AR Glasses in Home Environments}

\author{Shuning Zhang}
\orcid{0000-0002-4145-117X}
\authornotemark[1]
\affiliation{%
  \institution{Tsinghua University}
  \city{Beijing}
  \country{China}
}
\email{zsn23@mails.tsinghua.edu.cn}

\author{Qucheng Zang}
\orcid{0009-0002-6114-4940}
\authornote{Equal contribution.}
\affiliation{%
  \institution{China Academy of Art}
  \city{Hangzhou}
  \country{China}
}
\email{3190602072@caa.edu.cn}

\author{Yongquan `Owen' Hu}
\orcid{0000-0003-1315-8969}
\affiliation{%
  \institution{National University of Singapore}
  \city{Singapore}
  \country{Singapore}
}
\email{yongquanhu.work@gmail.com}

\author{Jiachen Du}
\orcid{0000-0001-6444-9112}
\affiliation{%
  \institution{Tsinghua University}
  \city{Beijing}
  \country{China}
}
\email{djc23@mails.tsinghua.edu.cn}

\author{Xueyang Wang}
\orcid{0000-0002-9797-9491}
\affiliation{
    \department{Institute for Network Sciences and Cyberspace}
    \institution{Tsinghua University}
    \city{Beijing}
    \country{China}
}
\email{wang-xy22@mails.tsinghua.edu.cn}

\author{Yan Kong}
\orcid{0000-0002-4187-3559}
\affiliation{
    \institution{Tsinghua University}
    \city{Beijing}
    \country{China}
}
\email{ky21@mails.tsinghua.edu.cn}

\author{Xinyi Fu}
\orcid{0000-0001-6927-0111}
\affiliation{
    \institution{The Future Laboratory, Tsinghua University}
    \city{Beijing}
    \country{China}
}
\email{fuxy@mail.tsinghua.edu.cn}

\author{Suranga Nanayakkara}
\orcid{0000-0001-7441-5493}
\affiliation{%
  \institution{Augmented Human Lab} 
  \institution{National University of Singapore} 
  \city{Singapore}
  \country{Singapore} 
}
\email{suranga@ahlab.org}

\author{Xin Yi}
\orcid{0000-0001-8041-7962}
\authornote{Corresponding author.}
\affiliation{
    \institution{Tsinghua University}
    \city{Beijing}
    \country{China}
}
\affiliation{
    \institution{Beijing Academy of Artificial Intelligence}
    \city{Beijing}
    \country{China}
}
\email{yixin@tsinghua.edu.cn}

\author{Hewu Li}
\orcid{0000-0002-6331-6542}
\affiliation{
    \institution{Tsinghua University}
    \city{Beijing}
    \country{China}
}
\email{lihewu@cernet.edu.cn}









\renewcommand{\shortauthors}{Zhang and Zang, et al.}


\begin{abstract}
    Always-on sensing of AI applications on AR glasses makes traditional permission techniques ill-suited for context-dependent visual data, especially within home environments. The home presents a highly challenging privacy context due to the high density of sensitive objects, and the frequent presence of non-consenting family members, and the intimate nature of daily routines, making it a critical focus area for scalable privacy control mechanisms. Existing fine-grained controls, while offering nuanced choices, are inefficient for managing multiple private objects. We propose VisGuardian, a fine-grained content-based visual permission technique for AR glasses. VisGuardian features a group-based control mechanism that enables users to efficiently manage permissions for multiple private objects. VisGuardian detects objects using YOLO and adopts a pre-classified schema to group them. By selecting a single object, users can efficiently obscure groups of related objects based on criteria including privacy sensitivity, object category, or spatial proximity. A technical evaluation shows VisGuardian achieves mAP50 of 0.6704 with only 14.0 ms latency and a 1.7\% increase in battery consumption per hour. Furthermore, a user study (N=24) comparing \proj{} to slider-based and object-based baselines found it to be significantly faster for setting permissions and was preferred by users for its efficiency, effectiveness, and ease of use. 
\end{abstract}


\begin{teaserfigure}
\centering
\includegraphics[width=1\textwidth]{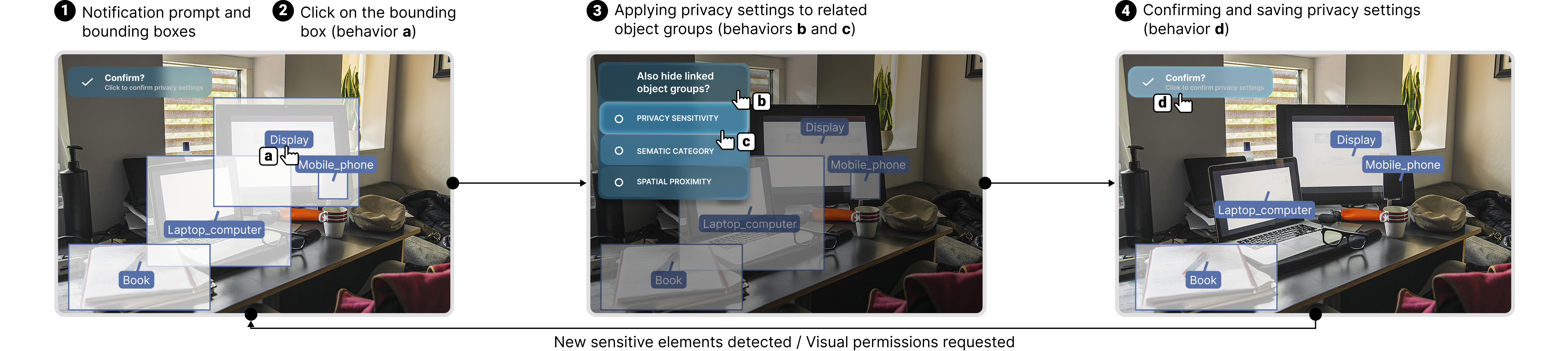}
\caption{\proj{} is a fine-grained permission technique for AR glasses, enabling group-based privacy management with minimal task disruption. (1) The reminding window pops up when sensitive objects of new classes are detected or when an app requests visual permissions. The streaming continues, and sensitive objects are highlighted with bounding boxes, (2) users can directly click the bounding box of different identified sensitive objects to hide or reveal them. When selected, (3) the interface pops up with the choices to obscure or reveal similar objects (including three choices based on: privacy sensitivity, object type, or spatial proximity). (4) After hiding the objects, \proj{} remembers the users' choice until new objects emerge or the user wants to manually change the settings.}
\label{fig:teaser}
\end{teaserfigure}

\maketitle

\section{Introduction}

The landscape of head-worn wearables is rapidly evolving towards a converging paradigm of Artificial Intelligence (AI) applications on Augmented Reality (AR). While early market distinguished between ``smart glasses'' (e.g., Ray-Ban Meta\footnote{https://www.meta.com/ai-glasses/ray-ban-meta/}), which focused on AI assistants and capture, and ``AR headsets'' (e.g., HoloLens\footnote{https://learn.microsoft.com/en-us/hololens/hololens-commercial-features}), which focused on immersive display, recent industry innovations, exemplified by prototypes like Meta's Orion\footnote{https://about.fb.com/news/2024/09/introducing-orion-our-first-true-augmented-reality-glasses/}, point towards a unified future: wearable glasses that integrate \textbf{always-on visual sensors} with \textbf{spatial AR displays}~\cite{rao2020investigating,suzuki2025everyday}. These devices\footnote{We used AR glasses to denote these devices in the latter sections of the paper.} leveraged advanced AI to proactively sense and interpret the user's environment, offering an always-on interface for visual understanding~\cite{cai2025aiget,zhao2024heads}. 

However, the persistent sensing capability introduces profound privacy risks, which are particularly acute within home environments. Unlike public or workplace settings, the home is a highly sensitive environment, a space for intimate family moments~\cite{zeng2019understanding}, private routines~\cite{chiang2020exploring}, and sensitive personal data~\cite{lenhart2023you}. Furthermore, the frequent co-presence of bystanders~\cite{marky2020don} (e.g., family members, guests) complicates the ethics and logistics of consent for continuous data collection. The home, therefore, represents a critical high-risk context that demands robust visual privacy controls.

The privacy risks extend beyond inadvertent data leakage. \textbf{A primary threat that motivates our work emerges from semi-trusted applications that provide services to the user but may extract and leak their private information.} This information could also be exploited for user profiling or targeted advertising, thereby compromising user privacy. Prior research has demonstrated that sensitive attributes, such as a user's gender~\cite{zhang2025pervasive,zhang2025through}, behavioral characteristics~\cite{berkovsky2019detecting,hoppe2018eye}, and emotional state~\cite{hickson2019eyemotion}, can be inferred in real time from the data captured by these visual sensors. For example, continuous sensing via the front camera could track a user's dietary habits from what they eat for breakfast, infer their emotional state from their facial expressions during a video call with a family member, or reveal their lifestyle habits from the books and documents on their desk.

While permission systems are the standard data control mechanism for AI applications on AR glasses~\cite{AndroidPermissions2022,OculusAndroid10Target}, their design is inadequate for the continuous data collection paradigm introduced by these devices~\cite{abraham2024you}. These sensors create a heightened privacy-utility tension, where the binary ``all-or-nothing'' nature of permissions forces an untenable choice~\cite{felt2011android,kelley2012conundrum}, a decision further complicated by users' poor comprehension of the implications~\cite{lin2012expectation,kelley2013privacy,johnstone2022virtuality}. 

Prior work has attempted to provide granular control~\cite{abraham2024you,venugopalan2024aragorn}. One stream of research focuses on multi-level settings~\cite{google_android_2022,abraham2024you,nair2023going} to move beyond simple binary choices. While these methods, such as privacy sliders, acknowledge the privacy-utility trade-off, they often lack the expressiveness to manage specific contextual objects and can impose a high cognitive load on users. Another stream of research involves automatic control paradigms~\cite{roesner2014world,steil2019privaceye} that configure protection policies or shutter cameras without user intervention. However, these paradigms did not consider users' preferences regarding privacy-utility trade-offs, and may compromise the task utility through shuttering the whole camera~\cite{steil2019privaceye}. The third stream enables direct data access control~\cite{aditya2016pic,raval2016you,jana2013enabling,lee2024priviaware,venugopalan2024aragorn}. These approaches offer great precision, evolving from manual policy configuration~\cite{lee2024priviaware} to automatic detection based on user selections~\cite{venugopalan2024aragorn}. However, they often require users to manage permissions on an object-by-object basis. This approach, especially when systems are designed around single salient objects such as credit cards~\cite{venugopalan2024aragorn}, does not scale well to complex scenarios, leading to high interaction costs. Therefore, a critical research gap exists for a technique that provides \textbf{fine-grained, content-based control through a low-cost, scalable interaction metaphor.} 




To address this gap, we present \proj{}, a content-based permission technique that introduces a group-based control mechanism to streamline privacy management for the data captured by the front cameras of AR glasses. Adhering to the principle of privacy-by-design~\cite{cavoukian2009privacy}, \proj{} first automatically detects and sanitizes sensitive objects (e.g., documents and screens) by default. To minimize the interaction cost of subsequent user adjustments, \proj{} allows users to select a single detected instance and apply permission settings across an entire group of semantically or spatially related items.. These groups are pre-defined along attributes such as privacy sensitivity, object category, or spatial proximity. To facilitate control, \proj{} renders in-situ semi-transparent overlays on detected objects, enabling users to anonymize or de-anonymize by directly clicking onthe detected objects. We envision \proj{} as a system-level service designed to enforce unified privacy controls across diverse applications.

Technically, \proj{} modifies the captured real-time video stream to sanitize salient objects. We chose YOLOv10~\cite{wang2024YOLOv10}, which is both efficient and effective, and fine-tuned this model on a hybrid dataset containing private objects such as people, driver's licenses, and underwear. The dataset was selected from a combination of LVIS~\cite{gupta2019lvis} and COCO train set~\cite{lin2014microsoft, lin2014modeling}, focusing on privacy-related objects. The sanitization process was implemented by visually occluding the detected objects. This approach was chosen for its computational efficiency, which is critical for real-time processing on AR glasses, and because prior work indicates that users find direct occlusion clearer and less ambiguous than partial blurring techniques~\cite{vishwamitra2017blur, li2018human}. \proj{} sanitizes all sensitive objects by default and remembers the user's settings for each category within an application. In a technical evaluation on combination of COCO and LVIS test sets, \proj{} showed comparable detection precision (0.6704) to other candidate architectures such as Faster RCNN (0.5560), YOLO-World (0.4080), and Detectron (0.5701). Additionally, \proj{} operates with an on-device latency of 14.0 ms and maintains smooth operation at 4 FPS, resulting in only a 1.7\% increase in hourly battery consumption compared to the baseline.

Finally, we conducted a user study (N=24) comparing \proj{} with two representative baselines: a slider-based control technique~\cite{abraham2024you} and an object-selection-based control technique~\cite{venugopalan2024aragorn}. The study was set within a simulated home environment, reflecting typical usage of AR glasses~\cite{arakawa2023prism, arakawa2024prism, steil2019privaceye, mahmud2024actsonic}. It included \textbf{daily health monitoring}, \textbf{household and lifestyle management}, \textbf{social interaction}, and \textbf{multimodal learning and work assistance} scenarios to evaluate how \proj{} facilitates the privacy-utility trade-off in practice. We found that \proj{} enabled users to complete their tasks effectively, was significantly faster and easier for managing permissions, and received higher user satisfaction ratings. In summary, the contributions of this work are threefold:

$\bullet$ We conceptualized and designed \proj{}, a \textbf{group-based} fine-grained visual privacy control technique on AR glasses.

$\bullet$ We demonstrated the \textbf{on-device technical feasibility} of \proj{}, with real-time latency of 14.0ms and a 1.7\% increase in hourly battery consumption.

$\bullet$ We provide \textbf{empirical evidence} on how users balance privacy and utility, and reduce control efforts using a group-based fine-grained control technique.

\section{Background \& Related Work}

Privacy issues and permission systems for AR glasses are important topics in the Human-Computer Interaction (HCI) community~\cite{venugopalan2024aragorn,opaschi2020uncovering,o2023privacy}. To situate our work, we first synthesize research on privacy challenges and implications associated with AR glasses. We then elaborate on past permission system designs, especially summarizing advancements in automatic permission control fine-grained permission systems. Finally, we synthesize the privacy categorization works in home environments to develop a taxonomy for our paper.

\subsection{Privacy Challenges and Implications in AR Glasses}

The pervasive and always-on visual sensing capabilities of AR glasses introduce severe privacy risks that extend beyond the wearer~\cite{harborth2021evaluating,shi2021face} to bystanders~\cite{denning2014situ,yao2019privacy,koelle2015don,bhardwaj2024focus} and the surrounding environment~\cite{kress2014segmentation,harborth2021evaluating}. These risks are magnified by the rich semantic details inherent in high-fidelity visual data, which can inadvertently expose sensitive information ranging from personal attributes and preferences~\cite{bye2019ethical,kress2014segmentation} to data captured via gaze cues~\cite{liebling2014privacy,wenzlaff2016video}, often without the awareness of those being recorded~\cite{koelle2018your,eiband2017understanding}. Expert insights further warn that such detailed real-time XR data facilitates not only personal identification but also the prediction of private conditions, potentially trapping users in polarized ``filter bubbles''~\cite{abraham2022implications}. \textbf{Because these privacy threats stem primarily from the captured content itself, addressing them requires mechanisms that operate effectively at the object level rather than merely controlling the sensor hardware.}

Furthermore, the privacy expectations for AR are highly context-dependent and granular. Empirical research has classified AR sensing privacy by factors such as nudity, data quantification risks, and confidentiality~\cite{abraham2024don}. Gallardo et al.~\cite{gallardo2023speculative} further cataloged user discomfort regarding specific data types, such as facial images and voiceprints, noting that users' privacy preferences vary across different sensitive inputs. This complexity implies that a simple ``camera-off'' switch may compromise the device's utility, creating a tension between privacy and functionality. \textbf{Consequently, static or binary permission models are insufficient. Instead, a dynamic, fine-grained control mechanism is required to allow users to navigate these complex trade-offs by selectively managing specific sensitive objects.}

To mitigate these threats, prior research has explored various countermeasures. Initial approaches focused on high-level controls, such as context-based camera controls~\cite{jung2014courteous,steil2019privaceye,shrestha2017offensive} or abstracting visual data into semantic tags~\cite{aiordachioae2019life,aiorduachioae2020aggregating}. However, they often fall short of providing users with fine-grained, content-based control in real time. More recent work has investigated using user-defined gestures for smart glasses' privacy protection~\cite{zhang2025through}, or copilot agents for managing photo sharing~\cite{monteiro2025imago}. While they advanced user's control, these methods do not fully address the continuous and dynamic nature of visual sensors on AR glasses. \textbf{This highlights the lack of an interaction metaphor that combines the precision of fine-grained control with the efficiency required for daily use, a gap our work resolves.}


\subsection{Privacy Permission Designs and Systems}\label{sec:rw_permission}

As AR devices evolve from specialized tools to always-on consumer products, the paradigm for privacy control has shifted. Traditional static permission models (e.g., install-time permission)~\cite{felt2012ask,felt2012android} and broad contextual cues (e.g., location or activity)~\cite{baarslag2016negotiation,wijesekera2018contextualizing,fu2014general} are increasingly insufficient for the continuous, visual nature of AR~\cite{roesner2014security,gallardo2023speculative}. AI applications introduced further complexity to this landscape, where the data usage were highly unpredictable. In response to this evolving landscape, past literature has focused on designing granular, dynamic privacy systems, which we categorize into three approaches: \textbf{automated context-aware protection}, \textbf{policy-driven frameworks} and \textbf{interaction-based control}.

\textbf{Automated Context-aware Protection} To minimize user burden, prior research has explored on automating privacy protection through computer vision (CV) and sensor fusion. Systems like \textit{PrivacEye}~\cite{steil2019privaceye} and work by Jung et al.~\cite{jung2014courteous} utilize eye tracking or thermal features to automatically disable cameras. \textbf{However, in home environments, sensitive objects often coexist with task-essential elements. Completely disabling the camera is therefore unsuitable, as it hinders task completion.} For lifelogging or in shared spaces, automated techniques have been proposed to detect and blur screens~\cite{korayem2016enhancing}, bystanders~\cite{hasan2020automatically,corbett2023bystandar} or faces~\cite{ye2014negative}. \textbf{While these techniques informed our approach, they target specific object category, offering limited exploration into the multiple objects' detection and interactive control in home environments.} Wijesekera et al.~\cite{wijesekera2017feasibility,wijesekera2018contextualizing} advanced dynamic permission prediction and adaptation based on user preferences. \textbf{However, their context-aware adaptations operates primarily at the sensor level, under-exploring the need for contend-based control in continuously visual streams.}

\textbf{Policy-Driven Frameworks} Policy-driven frameworks seek to structure control through pre-defined rules. The foundational work on \textit{World-Driven Access Control} by Roesner et al.~\cite{roesner2014world} linked access policies to physical objects via markers (e.g., via QR codes or visual markers), allowing objects to ``declare'' their own privacy policies. \textbf{Yet, relying on physical markers is impractical in dynamic home settings due to the high setup cost and the difficulty of tagging moving objects (e.g., people).} Similar work by Jana et al.~\cite{jana2013enabling} proposed abstracting sensor data (e.g., skeletons). \textbf{However, such abstractions limit utility when AI applications require reasoning based on contextual details.} Frameworks like Erebus~\cite{kim2023erebus} enable fine-grained control through developer-defined, user-modifiable policies. \textbf{Nevertheless, the requisite need for users to modify permissions via a companion app is unnatural for the AR environment, especially during continuous AI application usage. The black-box nature of AI applications also makes it difficult for developers to pre-define object-level access requirements.} Recent work has also explored multi-user scenarios. Rajaram et al. explored negotiation interfaces~\cite{rajaram2023eliciting} and equilibrium models~\cite{rajaram2025privacy} to balance conflicting privacy needs in shared AR spaces. \textbf{Different from the multi-user focus, we target at the object-level trade-offs between privacy and utility especially in AI application usage scenarios.}

\begin{table*}[t]
\centering
\caption{Comparison of visual privacy control techniques in related literature. For detection object, if it is written like ``person only'' or ``screen only'', they are targeted only to detect and redact person or screen. Otherwise, if it is said as ``object-level'', the permission or control is meant to be applied at the object level.}
\label{tab:comparison}
\resizebox{\textwidth}{!}{
\begin{tabular}{m{3.2cm} m{0.8cm} m{2.1cm} m{3cm} m{2.3cm} m{2.5cm} m{2.6cm} m{2cm} }
\toprule
\textbf{Paper} & \textbf{Year} & \textbf{User-Centric} & \textbf{D etection \& Interaction Pattern} & \textbf{Detect Objects \& Granularity} & \textbf{Device} & \textbf{Scenario} & \textbf{On-Device} \\
\midrule

Jana et al. \cite{jana2013enabling} & 2013 & Yes & Policy-based & Object & AR Glasses & General & Yes \\

Ye et al.~\cite{ye2014negative} & 2014 & Yes & Policy-based & Person only & Smart Glasses & Lifelogging & Not specified \\

Roesner et al. \cite{roesner2014world} & 2014 & No & World-driven & Object & Camera & General & Yes  \\

Jung et al. \cite{jung2014courteous} & 2014 & No & Temperature & Person only & Camera & Social & Yes \\

Korayem et al. \cite{korayem2016enhancing} & 2016 & No & Visual feature & Screen only & Wearables & Lifelogging & Yes \\

Raval et al. \cite{raval2016you} & 2016 & Yes & Object & Object & Mobile & General & Yes \\

Steil et al. \cite{steil2019privaceye} & 2019 & No & Visual feature & Sensor & Smart Glasses & General & Yes \\

Aiordăchioae et al. \cite{aiordachioae2019life} & 2019 & Yes & Visual feature & Sensor & Smart Glasses & Lifelogging & No \\

Hasan et al. \cite{hasan2020automatically} & 2020 & No & Visual feature & Sensor & Camera & Photos & Not specified \\

Corbett et al. \cite{corbett2023bystandar} & 2023 & No & Visual/Audio/Gaze & Bystander & AR Glasses & General & Yes \\ \midrule 

Kim et al. \cite{kim2023erebus} & 2023 & Yes & Policy-based & Object & AR Glasses & General & Yes \\

Nair et al. \cite{nair2023going} & 2023 & Yes & Policy-based & Person only & VR Glasses & General & Yes \\


Venugopalan et al. \cite{venugopalan2024aragorn} & 2024 & Yes & Object & Object & Mobile & General & Partial \\

Abraham et al. \cite{abraham2024you} & 2024 & Yes & Slider-based & Sensor & AR Glasses & General & Not Specified \\


Monteiro et al. \cite{monteiro2025imago} & 2025 & Yes & Object & Object & Mobile Camera & Image Share & No \\

Rajaram et al. \cite{rajaram2025privacy} & 2025 & Yes & Negotiation & Person only & AR Glasses & Multi-user & Not Specified \\
\midrule 

\textbf{VisGuardian} & -- & \textbf{Yes} & \textbf{Group-based} & \textbf{Object} & \textbf{AR Glasses} & \textbf{Home} & \textbf{Yes} \\
\bottomrule
\end{tabular}
} 
\renewcommand{\arraystretch}{1.0} 
\end{table*}

\textbf{Interaction-Based Control} To bridge the gap between automation and rigid policies, recent HCI research has explored metaphors for direct user control. This exploration centers on two interaction patterns: abstract level control and instance-level control. For \textbf{abstract level control}, systems simplify choices by offering high-level abstractions, such as privacy sliders to balance privacy and utility~\cite{nair2023going,abraham2024you} or defining sensitive context categories~\cite{abraham2024don}. While efficient, these abstract controls often lack the granularity to distinguish between specific objects within the same scene (e.g., blocking a \textit{confidential document} while sharing a \textit{book}). For \textbf{instance-level control}, the most granular approach allows users to explicitly mark specific regions or objects. Prior work has explored this through various modalities: marking 2D/3D surfaces~\cite{raval2016you} or identifying objects for redaction~\cite{venugopalan2024aragorn,monteiro2025imago}. While the visual metaphor of ``bounding boxes'' or ``visual markers'' for control is established~\cite{roesner2014world,raval2016you,cheng2024user,ligman2024data}, previous work required users to manage privacy on a \textit{per-object} or \textit{per-pixel} basis. \textbf{This per-object interaction~\cite{venugopalan2024aragorn} faces limitations in home environments, where AI applications reason on multiple contextual objects, leading to high user control overhead.}

Therefore, there remains a lack of an interaction metaphor that combines the \textit{granularity} of object-level control~\cite{raval2016you,venugopalan2024aragorn} while improving \textit{efficiency}. \proj{} addresses this by introducing a \textbf{group-based} control mechanism, allowing users to leverage semantic grouping to manage multiple related objects simultaneously. 


\subsection{Role-based Access Control and Attribute-based Access Control}

Foundational research in access control began with the Role-Based Access Control (RBAC) model. Ferraiolo et al.~\cite{ferraiolo1999role,ferraiolo2001proposed} developed the standard for RBAC to unify authorization policies, facilitating scalable administration. In modern smart home contexts, researchers have adapted RBAC to handle dynamic risks, such as Abusini et al.'s risk-based model~\cite{abusini2024enhancing} and Ameer et al.'s hybrid role-centric models~\cite{ameer2022hybrid}. While RBAC provides administrative efficiency through role grouping, its role assignment is ill-suited for the dynamic usage context of AI-based applications in home. \proj{} adopts the RBAC's philosophy of efficiency by allowing users to manage groups of objects rather than individual instances. 

To address the flexibility limitations of RBAC, Attribute-Based Access Control (ABAC) was introduced to evaluate authorization based on subject, object and environmental attributes~\cite{hu2014guide,hu2014attribute}. Kuhn et al.~\cite{kuhn2010adding} and Yuan et al.~\cite{yuan2005attributed} demonstrated that combining RBAC with ABAC attributes substantially enhances scalability and flexibility for distributed applications. \proj{} operationalizes ABAC by translating attributes--specifically object category, privacy sensitivity, and spatial proximity--into a user-facing interaction metaphor, making complex attribute-based rules visually manageable for end-users.

Recent work has further extended ABAC to address specific privacy and context-aware challenges in IoT and cloud environments. Smari et al.~\cite{smari2014extended} and Ed et al.~\cite{ed2016formal} integrated trust and privacy requirements directly into ABAC models. Others have focused on dynamic environments, such as Kolter et al.'s user-defined privacy rules~\cite{kolter2007privacy}, Van et al.'s context-aware ``Break Glass'' architecture~\cite{van2020context}, and granular filtering frameworks for cloud and IoT proposed by Son et al.~\cite{son2019novel} and Bhatt et al.~\cite{bhatt2020abac}. Unlike these systems which focus on network-level  policy enforcement or automated filtering, \proj{} bridges the gap between application-wise privacy attributes and control, empowering users to negotiate these privacy-utility trade-offs.

\subsection{Privacy Categorization}\label{sec:rw_privacy_cate}

To manage visual privacy risks, research focused on defining and categorizing sensitive visual content. Li et al.~\cite{li2018human} synthesized prior work to propose foundational categories such as identity, nudity and impression management, later refining them~\cite{li2020towards} in the context of photo sharing to include granular concepts like irresponsible behavior and medically sensitive content. From a data-centric perspective, Orekondy et al.~\cite{orekondy2017towards} identified 67 categories of private information in visual datasets, later organizing them into broad classes of textual (e.g, names, addresses), visual (e.g., faces, license plates) and multimodal information (e.g., credit cards)~\cite{orekondy2018connecting}. 

Other frameworks have been developed for specific contexts, such as the privacy classification by Wang et al.~\cite{wang2023modeling} for Activities of Daily Living (ADLs) which distinguishes between biometric identification and societal information. Similarly, Yu et al.~\cite{yu2018leveraging} introduced a hierarchical framework categorizing content as personalized, adult and common. Collectively, these evolving framework highlight the diverse and context-dependent nature of sensitive content, which we operationalized as machine-detectable private objects (e.g., faces, ID cards, documents) for detection.

\section{Visual Private Objects' Categorization Framework In Home Environments}\label{sec:categorization}

To enable efficient and intuitive content-based privacy control, we propose a structured classification framework for sensitive visual information. This framework is theoretically grounded in established access control models. Specifically, we adopt the administrative efficiency philosophy of RBAC~\cite{ferraiolo1999role,ferraiolo2001proposed} by allowing users to manage groups of private objects rather than individual instances. Furthermore, to address the dynamic and contextual nature of augmented reality in home environments, our framework operationalizes ABAC~\cite{hu2014guide,hu2014attribute,yuan2005attributed} by translating three core attributes--privacy sensitivity, object category, and spatial proximity--into a user-manageable interaction metaphor.

\begin{table*}[!htbp]
\centering
\caption{Taxonomy of visual privacy objects and their risk categories. We mapped theoretical categorizations from prior work (representative objects) to the specific instances (evaluated instances), risk levels, and spatial contexts. Text in \textcolor{typeDigital}{\textbf{green}} denotes objects added compared to prior work.}
\label{tab:merged_privacy_objects}
\small
\resizebox{\textwidth}{!}{%
\begin{tblr}{
  colspec = {l l p{3.35cm} p{2.65cm} p{1.5cm} p{1.5cm} p{2cm}},
  row{1} = {font=\bfseries, bg=gray!15, m}, 
  column{1,2} = {font=\bfseries}, 
  vlines = {white}, 
  hlines = {0.5pt, gray!50}, 
  hline{1,2,Z} = {1pt, black}, 
  cell{1}{1,2} = {r=1,c=1}{}, 
  stretch = 0.95, 
}
Category & Subcategory & {Representative Objects} & {Evaluated Instances} & Risk & Space & Type \\

\SetCell[r=9]{m} Identity-Related & Human Face & Identifiable/Partial face~\cite{li2018human,orekondy2018connecting} & person & \SetCell{bg=riskHigh} High & \SetCell{bg=spatialPersonal} Personal & \SetCell{bg=typeMarker} Personal Marker \\
 & Body Parts & Fingerprints, Tattoos~\cite{orekondy2018connecting} & \textit{N/A} & - & - \\
 & \SetCell[r=3]{m} Clothing & \SetCell[r=3]{m} Uniforms, Logos~\cite{li2020towards,yu2018leveraging} & underwear & \SetCell{bg=riskHigh} High & \SetCell{bg=spatialPersonal} Personal & \SetCell{bg=typeClothes} Clothes \\
 & & & swimsuit & \SetCell{bg=riskMedium} Med & \SetCell{bg=spatialActivity} Activity & \SetCell{bg=typeClothes} Clothes \\
 & & & legging, pajamas, skirt & \SetCell{bg=riskLow} Low & \SetCell{bg=spatialBedroom} Bedroom & \SetCell{bg=typeClothes} Clothes \\
 & \SetCell[r=2]{m} Accessories & \SetCell[r=2]{m} Glasses, Watches~\cite{orekondy2017towards} & jewelry & \SetCell{bg=riskHigh} High & \SetCell{bg=spatialPersonal} Personal & \SetCell{bg=typeDigital} Digital \\
 & & & badge & \SetCell{bg=riskLow} Low & \SetCell{bg=spatialOffice} Office & \SetCell{bg=typeMarker} Personal Marker \\
 & \SetCell[r=2]{m} Documents & \SetCell[r=2]{m} Passports, ID Cards~\cite{orekondy2018connecting}, \newline Vehicle plates~\cite{li2020towards} & license plate & \SetCell{bg=riskMedium} Med & \SetCell{bg=spatialActivity} Activity & \SetCell{bg=typeUncat} Others \\ 
 & & & ID card, checkbook, \newline signed document & \SetCell{bg=riskMedium} Med & \SetCell{bg=spatialOffice} Office & \SetCell{bg=typeMarker} Personal Marker \\

\SetCell[r=2]{m} Location-Based & \SetCell[r=2]{m} Home Spaces & \SetCell[r=2]{m} Furniture, Bedrooms~\cite{li2020towards} & toilet & \SetCell{bg=riskMedium} Med &  \SetCell{bg=spatialBathroom} Bathroom & \SetCell{bg=typeSafety} Safety \\
 & & & file cabinet, book & \SetCell{bg=riskLow} Low &  \SetCell{bg=spatialOffice} Office & \SetCell{bg=typeAppendences} Appendences\\

\SetCell[r=4]{m} Physical Traits & \SetCell[r=2]{m} Health & \SetCell[r=2]{m} Medicine, Devices~\cite{li2018human} & medicine & \SetCell{bg=riskHigh} High &  \SetCell{bg=spatialLiving} Living & \SetCell{bg=typeUncat} Others \\
& & & wheelchair & \SetCell{bg=riskMedium} Medium & \SetCell{bg=spatialBathroom} Bath & \SetCell{bg=typeSafety} Safety \\
 & Facial Expr. & Crying, Forced smiles~\cite{li2018human} & \textit{Implicit in `person'} & - & - & -\\
 & Nudity & Partial nudity~\cite{li2020towards} & \textit{Implicit in `underwear'} & - & - & -\\

\SetCell[r=4]{m} Digital \& Textual & \SetCell[r=2]{m} Digital Devices & \SetCell[r=2]{m} Smart phone, laptop & \textcolor{typeDigital}{mobile phone} & \SetCell{bg=riskMedium} Med &  \SetCell{bg=spatialPersonal} Personal & \SetCell{bg=typeDigital} Digital \\
 & & & \textcolor{typeDigital}{laptop computer} & \SetCell{bg=riskMedium} Med &  \SetCell{bg=spatialActivity} Activity & \SetCell{bg=typeDigital} Digital \\
 & Identity Texts & Names, Emails, Phone~\cite{orekondy2017towards} & \textit{Content on screens} & \SetCell{bg=riskMedium} Med & - & - \\
 & Sensitive Docs & Receipts, Statements~\cite{li2020towards} & calendar & \SetCell{bg=riskMedium} Med &  \SetCell{bg=spatialOffice} Office & \SetCell{bg=typeUncat} Others \\

\SetCell[r=2]{m} Safety \& Social & Safety & Smoking, Drinking, \newline Weapons~\cite{li2020towards,orekondy2017towards} & gun, drunk & \SetCell{bg=riskMedium} Med &  \SetCell{bg=spatialActivity} Activity & \SetCell{bg=typeSafety} Safety \\
 & Valuables & Jewelry, Cash~\cite{orekondy2018connecting} & \textit{See `Accessories'} & - & - & - \\
\end{tblr}
}%
\end{table*}

We synthesized prior established taxonomies of visual privacy~\cite{orekondy2017towards,orekondy2018connecting,li2018human,wang2019systematic} to construct a foundation of detectable privacy objects (see Table~\ref{tab:merged_privacy_objects}). These are objects that can be computationally identified by detection models (e.g., YOLO~\cite{cheng2024yolo}). The initial list was derived from a literature review~\cite{li2018human,wang2023modeling,orekondy2018connecting,li2020towards,yu2018leveraging,orekondy2017towards}, with the addition of emerging, high-risk objects like smartphones and laptops. We acknowledge that some privacy concepts (e.g., ``drunk'') are actions or states, not physical objects. These are mapped to detectable object proxies with similar privacy implications (e.g., \textit{drunk} is represented by \textit{beer} or \textit{alcohol}, \textit{partial nudity} and \textit{tattoos} are mapped to \textit{person}). One author conducted this mapping, and the other authors checked and resolved potential disagreements. The complete list and their subsequent groupings are shown in Table~\ref{tab:merged_privacy_objects}. 

As prior work in HCI and privacy suggests that users' mental models of privacy are inherently contextual~\cite{wang2023modeling,oishi2020semantic}, our three core dimensions are designed to capture common contextual rules and align with user expectations for a group-based control mechanism. We defined the three dimensions for attribute-based control as follows:

$\bullet$ \textbf{Privacy Sensitivity}: This dimension categorizes objects based on their potential for harm of personal information disclosure~\cite{wang2023modeling}. It allows users to enforce broad authorization policies, such as ``hide all high-sensitivity items''. We adopted the base sensitivity rating scores from prior work~\cite{wang2023modeling}, with experimenters empirically determined the sensitivity of new or uncategorized objects through collaborative discussion.

$\bullet$ \textbf{Object Category}: This dimension groups objects based on conceptual or categorical similarity, aligning with how users understand and categorize their environment (e.g., ``hide all documents'' or ``hide all screens'')~\cite{li2018human,wang2023modeling,orekondy2018connecting}. For instance, all screen-like objects (\textit{laptop}, \textit{mobile phone}) are grouped together, as are all forms of identification (\textit{ID card}, \textit{passport}). In this work, objects correspond to the 7 object type groups detailed in Table~\ref{tab:merged_privacy_objects}.

$\bullet$ \textbf{Spatial Proximity}: This dimension groups objects based on their physical location or spatial context, enabling users to apply rules according to their immediate environment, such as ``hide objects in the office space.'' This is particularly relevant for AR systems due to their constant context changes. The five spatial groups are mapped in Table~\ref{tab:merged_privacy_objects}.

To categorize objects not covered by prior works~\cite{orekondy2017towards,orekondy2018connecting,wang2023modeling}, two authors collaboratively developed the classifications of all three dimensions (\textit{Privacy Sensitivity}, \textit{Object Category}, and \textit{Spatial Proximity}). The researchers independently coded the objects and then met to discuss and resolve discrepancies. Disagreements were resolved through iterative discussions.

\section{Design of \proj{}}

\subsection{Design Goals}
In AR, the continuous nature of visual sensing for AI applications requires content-based privacy management and real-time asynchronous user control. These distinct challenges motivate the following design goals:

\textbf{DG1: Content-based Permission Management} Since AI applications on AR glasses continuously track the user's environment, privacy permissions should be context-sensitive, focusing on the content being captured. Unlike conventional privacy settings that may only control access to data sources (e.g., sensors), content-based permissions should determine what types of visual information (e.g., faces, sensitive objects, or locations) can be processed, stored or shared. 

\textbf{DG2: Continuous and Non-intrusive Permission Feedback} Due to the always-on nature of visual sensing, users should receive continuous, non-intrusive and asynchronous feedback on their privacy settings. This feedback should indicate when specific content is being captured or processed, enabling users to modify permissions dynamically without interrupting the ongoing activity. This real-time awareness is different from traditional privacy permissions~\cite{felt2012android}, which offer static controls without granular, real-time awareness.

\textbf{DG3: Proactive Protective Design} AR glasses should feature proactive access control without needing users' repetitive actions. For example, with changing environments, the system should apply previous settings as defaults, while providing alerts about emerging private objects, complying with the privacy-by-design principle~\cite{langheinrich2001privacy}.

\textbf{DG4: Achieving Asynchronous Transparent Control} To avoid user fatigue, especially in continuous sensing scenarios, real-time control over visual data should be non-intrusive. Instead of requiring users to directly intervene in every moment of data capture, \proj{} should allow for broad, pre-configured privacy preferences. These preferences should operate asynchronously, automatically adapting to varying contexts, thereby reducing the need for constant user attention.

\subsection{Overview of \proj{}}

\proj{} is designed to enhance traditional sensor-based permissions by introducing a content-based permission framework, automatically suggesting group-based control. \proj{} also remembers users' choices, similar to runtime privacy notices~\cite{li2024we}. 

\begin{figure}
    \includegraphics[width=0.5\textwidth]{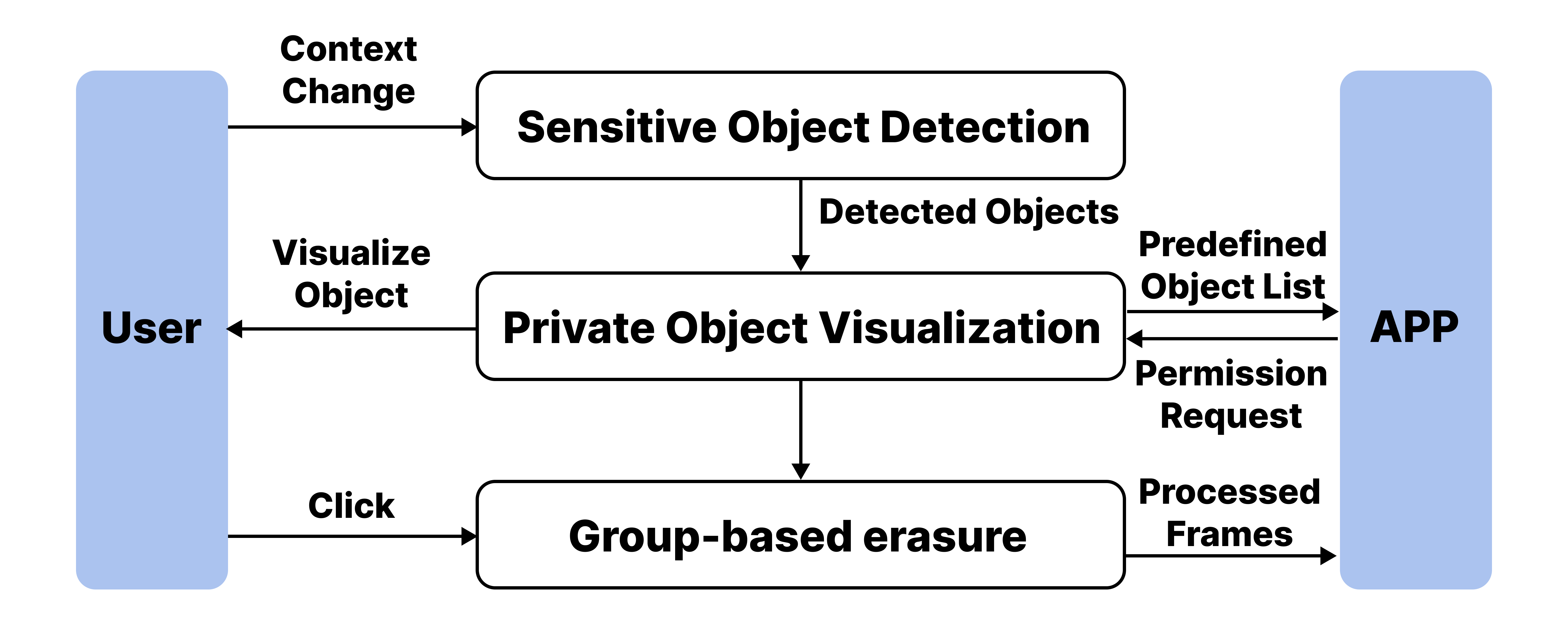}
    \caption{The system flow of \proj{}, where the processed video would be sent to the app upon users' group-based control. \proj{} checks for privacy-sensitive content before the app has access to a video stream.}
    \label{fig:algorithm_flow}
\end{figure}

As shown in Figures~\ref{fig:teaser} and~\ref{fig:algorithm_flow}, \proj{} employs an on-device object detection mechanism and adaptively groups objects based on the proposed privacy categorization framework. All detected privacy objects are sanitized initially, adhering to the privacy-by-design principle~\cite{cavoukian2009privacy, cavoukian2021privacy}. It allows users to actively choose to hide or reveal identified private objects before sharing them with AI-powered applications. In this context, the user's action to hide an object serves as a direct command for \proj{} to apply its sanitization process (i.e., occluding the object with an opaque overlay) on the specific private object in the video stream. It further prompts users with group-based selections to reduce the effort required for user-initiated modifications. When users manipulate objects, \proj{} automatically sanitizes the corresponding objects in the video stream in the background. The processed frames are shown to the users in real time and sent to the AI-empowered application. \proj{} also remembers users' past selections and will default to these remembered settings until new objects emerged or users change the setting. This approach ensures that users can efficiently protect sensitive visual information while maintaining a smooth task experience. 

\subsection{Interaction Design}

\proj{} features a minimalist design to enable real-time, continuous content-based permission management. The permission interface automatically pops up upon users putting on the AR glasses, allowing them to select from multiple permission settings (see Figure~\ref{fig:teaser} \textcircled{2}), similar to runtime privacy notices~\cite{schaub2015design}. We adopted a peripheral pop-up instead of direct pop-ups~\cite{muller2015increasing,nan2015uipicker} or screen overlays to provide a non-intrusive, asynchronous experience that does not interfere with the user's current view or task. Upon activation, different permission settings are previewed, with the initial selection causing the preview to overlay on the original video stream~\cite{prange2021priview}, clearly indicating the applied privacy protection.

While default settings are in place, users have different controls through simple interactions. Users can tap, point, or select individual or multiple detected objects. This will trigger a panel prompting users to select from one or multiple groups (i.e., \textit{privacy sensitivity}, \textit{object category} and \textit{spatial proximity} group), where users choose whether or not to hide other similar objects. Users can choose to erase the corresponding groups based on privacy sensitivity, object category, or spatial proximity, as outlined in Section \ref{sec:categorization}. For example, if a user selects to erase underwear, then objects with the same category (e.g., clothes), the objects with the same proximity range (e.g., those commonly appearing in bedrooms), and the objects with similar sensitivity, (e.g., person) will form three groups respectively. The user can click to automatically erase the objects in the whole group. We chose directly obscuring rather than only blurring, as prior work indicates that users find direct occlusion clearer, less ambiguous, and more trustworthy than partial blurring techniques~\cite{vishwamitra2017blur}.

When selected, the grouped objects are highlighted to emphasize the applied privacy protections~\cite{ishida2024designing}. Users can select one or more on the displayed panel. This design prioritizes simplicity, relying on basic interactions like tapping and pointing~\cite{yu2017tap}. The preview updates dynamically according to users' movements. However, if users select individual objects or object groups, the preview will pause briefly until the selection is complete. Once confirmed, \proj{} will update the view to reflect the selected privacy settings, ensuring synchronization with the user's viewpoint (see Figure~\ref{fig:teaser} \textcircled{4}). Upon finalizing their selection, users can continue with other tasks, while the interface minimize to a peripheral bar. The interface will reappear under the following conditions: (1) \proj{} detects a new object that conflicts with the current settings or requires user input for further control, or (2) users proactively click on the bar to adjust their settings.
 
\subsection{Interface Design}

\proj{}'s interface mimics runtime permission choices~\cite{felt2012android}, thereby reducing learning cost. The interface emphasizes clarity and ease of interaction. Upon activation of the camera sensor, \proj{} presents a preview of the recorded interface with all private objects anonymized, where users can interact to de-anonymize objects. These anonymized objects are dynamically rendered as overlays on the preview, enabling real-time feedback. Changes made through users' interactions are immediately reflected in the preview as overlays or not overlays, providing real-time visual feedback. 

\section{Implementation of \proj{}}\label{sec:implementation}

We used the Mixed Reality Toolkit (MRTK)~\cite{mrtk_unity}, and MRTK-UI to develop \proj{}'s interfaces. Specifically, the group-selection panel was implemented using MRTK's \verb|PressableButton| prefabs and scripted to always face the user, ensuring clear visibility and accessibility during interaction. We used Universal Windows Platform (UWP) for controlling the permissions on the specific AR device, which provides access to system-level APIs, facilitating the control of camera recordings. We used the Sentis~\cite{unity_sentis} library and the Burst compiler for neural network-related processing. \proj{} consists of two modules: it first detects salient objects continuously, and then adaptively adopts sanitization algorithms on the video stream.

\subsection{Continuous Saliency Object Detection and Understanding in Videos}\label{sec:saliency_object}

Numerous models have the capability to perform object detection and understanding~\cite{wang2024YOLOv10,ren2016faster}. Since the focus of this paper is on content-based permission setting rather than benchmarking object detection algorithms, we prioritized the \textbf{latency and size} of the models. Because sensitive content was already categorized in previous works~\cite{wang2023modeling,wu2020privacy} and in this paper (Section~\ref{sec:categorization}), we considered both open-set~\cite{cheng2024yolo} and closed-set~\cite{ren2016faster,wang2024YOLOv10} detection models. Among open-set detection models, YOLO-World has the lowest latency and highest accuracy. Among closed-set detection models, YOLOv10 has the least latency and highest accuracy. Given that YOLO-World is several times larger than YOLOv10 and that our visual privacy objects are clearly scoped, we chose YOLOv10. As YOLOv10 was previously trained on the COCO dataset~\cite{lin2014microsoft}, this model was not curated for privacy detection. To enable it to detect and understand privacy objects, we fine-tuned YOLOv10 on COCO and LVIS dataset~\cite{gupta2019lvis}, which contained images with all the private object categories needed for this paper. We selected the labels according to the objects listed in Table~\ref{tab:merged_privacy_objects}. The mapping from object category to labels in the dataset were manually set by two authors, with intermittent discussions to resolve discrepancy (see  Appendix~\ref{sec:distribution} for the label details and distribution). We utilized all the training data with specific private objects and the corresponding labels for training. We set the hyperparameters as in Table~\ref{tab:hyp_params}. The comparison of YOLOv10 and other models in the aspect of efficiency and accuracy was verified in Section~\ref{sec:technical}.

\begin{table}[h!]
\centering
\caption{Hyperparameter settings for our selected technique.}
\label{tab:hyp_params}
\resizebox{0.5\textwidth}{!}{
    \begin{tabular}{ccccc} 
    \toprule
    \textbf{lr0} & \textbf{lrf} & \textbf{momentum} & \textbf{weight\_decay} & \textbf{warmup\_epochs} \\
    0.01 & 0.01 & 0.937 & 0.0005 & 3.0 \\
    \midrule
    \textbf{warmup\_mom.} & \textbf{warmup\_bias\_lr} & \textbf{box} & \textbf{cls} & \textbf{dfl} \\
    0.8 & 0.1 & 7.5 & 0.5 & 1.5 \\
    \bottomrule
    \end{tabular}
}
\end{table}

\subsection{Sanitization Implementation}

To ensure privacy, our sanitization process is based on visual occlusion. This process is applied after object detection, targeting the bounding boxes that enclose salient objects (see Section~\ref{sec:saliency_object}). The occlusion is achieved by dynamically generating a primitive \verb|Quad| mesh in the 3D space that precisely covers the area of the detected object from the user's perspective. The four corner points of the object's bounding box are first projected into the world space. A \verb|Quad| is then created at the center of these points. Its scale is adjusted to match the width and height calculated from the corner positions, ensuring it fully covers the sensitive object.

To make the occlusion visually informative yet non-distracting, the material applied to the \verb|Quad| uses a non-transparent shader. We extract the texture of the detected object from the video frame and apply it to the \verb|Quad|. This technique effectively freezes the object's appearance at the moment of detection, hiding any subsequent changes or sensitive details while still providing context to the user about what is being hidden. This method is computationally lightweight compared to real-time blurring shaders, making it ideal for resource-constrained devices like ARs. 
 
\section{Technical Evaluation}\label{sec:technical}

We evaluated system efficiency and accuracy using the open-source LVIS~\cite{gupta2019lvis} and COCO~\cite{lin2014microsoft} datasets. These were selected as they are common benchmarks~\cite{venugopalan2024aragorn}, contained the specific privacy objects relevant to this paper, and provided sufficient data for model training. This precluded the need for larger datasets like Object365~\cite{shao2019objects365} or specialized ones such as PA-HMDB51~\cite{wu2020privacy}.

We tested several model architectures. We included a Faster R-CNN based model utilizing a ResNet50 backbone, chosen over VGG to reduce parameter size and over MobileNet to enhance accuracy. Furthermore, various models from the YOLO family, including advanced versions like YOLOv10, were evaluated across different scales (nano to medium) to assess performance trade-offs. For comparative purposes, we also tested an open-world detection model.

For training, we empirically optimized learning rate using line search and set all hyperparameters as default. To ensure result robustness, each model variant was trained three times independently using different random seeds. Evaluation encompassed accuracy metrics (precision, recall, mAP@IoU=0.5, mAP@IoU=[0.5,0.95]) and efficiency metrics (model size, latency). Inference latency was measured on a server equipped with an Nvidia A10 GPU (24GB). Table~\ref{tab:performance_table} shows the results of this evaluation.

\begin{table}[h!]
\centering
\caption{Performance of different models. The latency was measured on the server. Model size is reported in million parameters (M). mAP50-95 denotes mAP@IoU=[0.5, 0.95]. mAP50 denotes mAP@IoU=0.5.}
\label{tab:performance_table} 
\resizebox{0.5\textwidth}{!}{
    \begin{tabular}{lcccccc}
    \toprule
    \textbf{Model} & \textbf{Precision} & \textbf{Recall} & \textbf{mAP50} & \textbf{mAP50-95} & \textbf{Latency (ms)} & \textbf{Model Size (M)} \\
    \midrule
    FasterRCNN (ResNet50) & 0.6501 & 0.5570 & 0.6122 & 0.4398 & 59.00 & 42.0 \\ 
    Yolov8s-worldv2 & 0.6658 & 0.5251 & 0.5608 & 0.4406 & 26.50 & 12.8 \\ 
    Yolov10n   & 0.6704 & 0.5360 & 0.5987 & 0.4455 & 4.38 & 2.3 \\ 
    Yolov10s   & 0.6914 & 0.6089 & 0.6641 & 0.5067 & 19.16 & 7.2 \\ 
    Yolov10m   & 0.7120 & 0.6095 & 0.6734 & 0.5171 & 22.66 & 15.4 \\ 
    \bottomrule
    \end{tabular}
}
\end{table}

The selection process prioritizes efficiency and detection accuracy. Based on this, the Faster R-CNN (ResNet50) exhibits both excessive latency (59.00 ms) and a prohibitive model size (42.0 M), rendering it impractical. Similarly, YOLOv10m, despite achieving the highest mAP scores, presents challenges with its large size (15.4 M) and relatively higher latency (22.66 ms) compared to smaller YOLO variants. The YOLOv8s-worldv2 model, while smaller than Faster R-CNN or YOLOv10m, has a substantially lower mAP50 accuracy (0.5608), less competitive latency (26.50 ms) and size (12.8 M) compared to YOLOv10n.

The decision consequently focuses on YOLOv10n versus YOLOv10s. YOLOv10s offers a notable improvement in accuracy (e.g., mAP50: 0.6641 vs 0.5987). However, this comes with a substantially larger model size (7.2 M vs 2.3 M). In contrast, YOLOv10n provides the lowest latency (4.38 ms) and smallest model size (2.3 M), which is crucial for the interaction on AR devices. While its accuracy is slightly lower than YOLOv10s, it remains competitive and robust. \textbf{Therefore, we selected YOLOv10n, as it presented the optimal trade-off, balancing strong performance with the minimal resources required for deployment.}

\section{User Evaluation in Real-World Scenarios with AR Glasses}
\label{sec:user_evaluation}

We conducted a study to evaluate \proj{} against representative fine-grained control methods. We hypothesized that \proj{}'s group-based mechanism is more efficient (H1), requiring less time and fewer interactions to configure, and would be perceived as more usable (H2), resulting in lower workload and higher user satisfaction than baseline methods.

\subsection{Participants and Apparatus}

\begin{figure}
    \centering
    \includegraphics[width=0.9\linewidth]{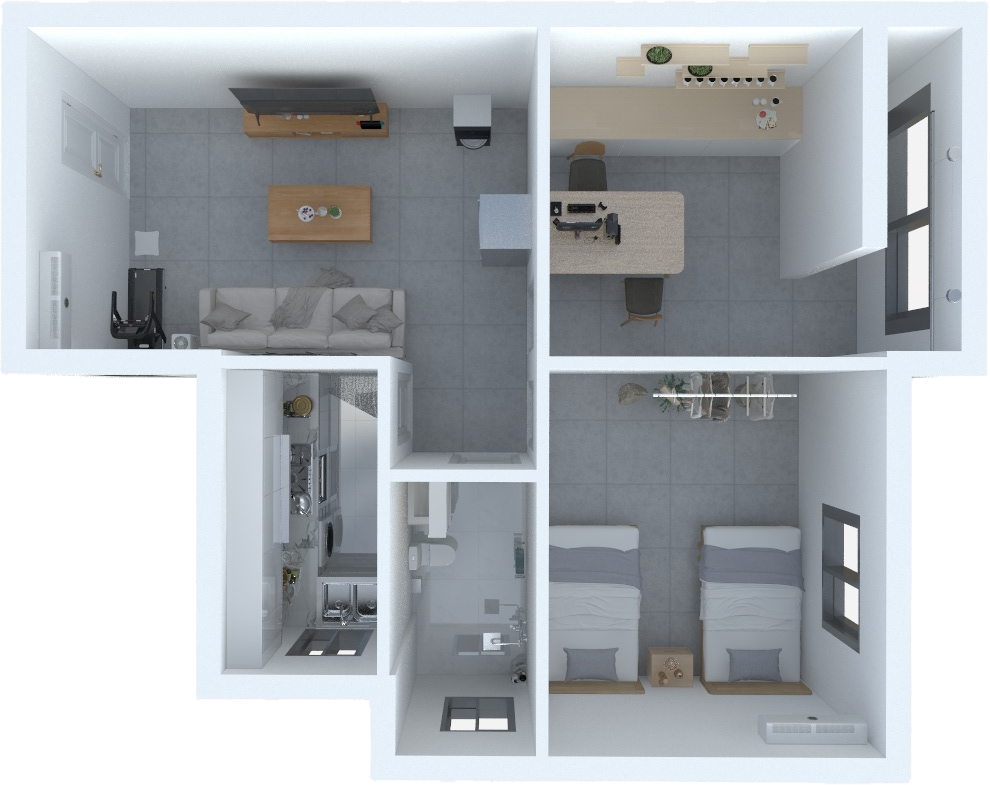}
    \caption{The home environment where we conducted the experiment.}
    \label{fig:environment}
\end{figure}

This IRB-approved study recruited 24 participants (13 male, 11 female) with a mean age of 24.5 (SD=4.6, age ranging from 18 to 35) by distributing the questionnaire on Wechat and the Xiaohongshu (Redbook). We adopted a Hololens 2 for the experiment. We carried out the experiment in a decorated home environment with a bedroom, study, living room, kitchen, and bathrooms (see Figure~\ref{fig:environment}), with preset privacy elements to avoid ethical concerns caused by users bringing their own private objects. Each participant was compensated 100 RMB, according to the local wage standard.

\subsection{Experiment Design}

To evaluate the efficiency of \proj{}'s group-based mechanism, we conducted a within-subjects study comparing it against two representative user-centric baselines: \textit{Object-based} (derived from Aragorn~\cite{venugopalan2024aragorn}) and \textit{Slider-based} (derived from Abraham et al.~\cite{abraham2024you}). These were selected for the specific problem of fine-grained, user-initiated management. While these designs primarily address data fidelity and availability, our study adapts their mechanisms (i.e., individual object selection and slider-based control) to the specific context of preventing the capture of perceived private objects. We excluded other paradigms for the following reasons: (1) \textbf{Binary controls}~\cite{jackson2018addressing,lin2014modeling} were excluded due to insufficient granularity for trade-off decisions, (2) \textbf{Policy-based systems}~\cite{roesner2014world,kim2023erebus} were excluded as configuring policies for each object causes high cost for users, and setting policies by experimenters omit the personalized privacy-utility trade-offs, (3) \textbf{Sensor-level automatic systems}~\cite{steil2019privaceye} were excluded as it did not feature object-level control for privacy-utility trade-offs, (4) \textbf{Object-detection systems}~\cite{jana2013enabling,hasan2020automatically,corbett2023bystandar,ye2014negative} were excluded as they scoped to detect humans or specific objects rather than general household objects, and (5) \textbf{RBAC/ABAC}~\cite{son2019novel,bhatt2020abac} frameworks were excluded as they focus on policy enforcement architectures rather than interaction metaphors. The tested conditions were:

\textbf{\proj{}}: We implement \proj{} as described in Section~\ref{sec:implementation}.

\textbf{Object-based}: To represent the paradigm of object manipulation, we designed a baseline conceptually derived from systems like Aragorn~\cite{venugopalan2024aragorn}. While Aragorn focused on managing one salient object, we enhanced this baseline to support the detection and selection of multiple objects simultaneously. In this condition, users could permit any number of objects but had to do so one by one. This facilitates the evaluation of \proj{}'s grouping mechanism.

\textbf{Slider-based}: We implemented a baseline inspired by the multi-level control paradigm, such as the `slider w/ image' setting in Abraham et al ~\cite{abraham2024you}. The relative sensitivity of the objects were determined following the survey in previous work~\cite{wang2023modeling}. The leftmost point corresponded to considering most about privacy (i.e., blurring all private objects) and the rightmost point corresponded to considering least about privacy (i.e., only blurring the most private objects)\footnote{See Appendix~\ref{app:mapping} for detailed descriptions of the mapping.}. This approach contrasts with object-based methods by abstracting control from direct object manipulation.

\begin{figure*}[!htbp]
    \subfloat[Object-based.]{
        \includegraphics[height=0.22\textwidth]{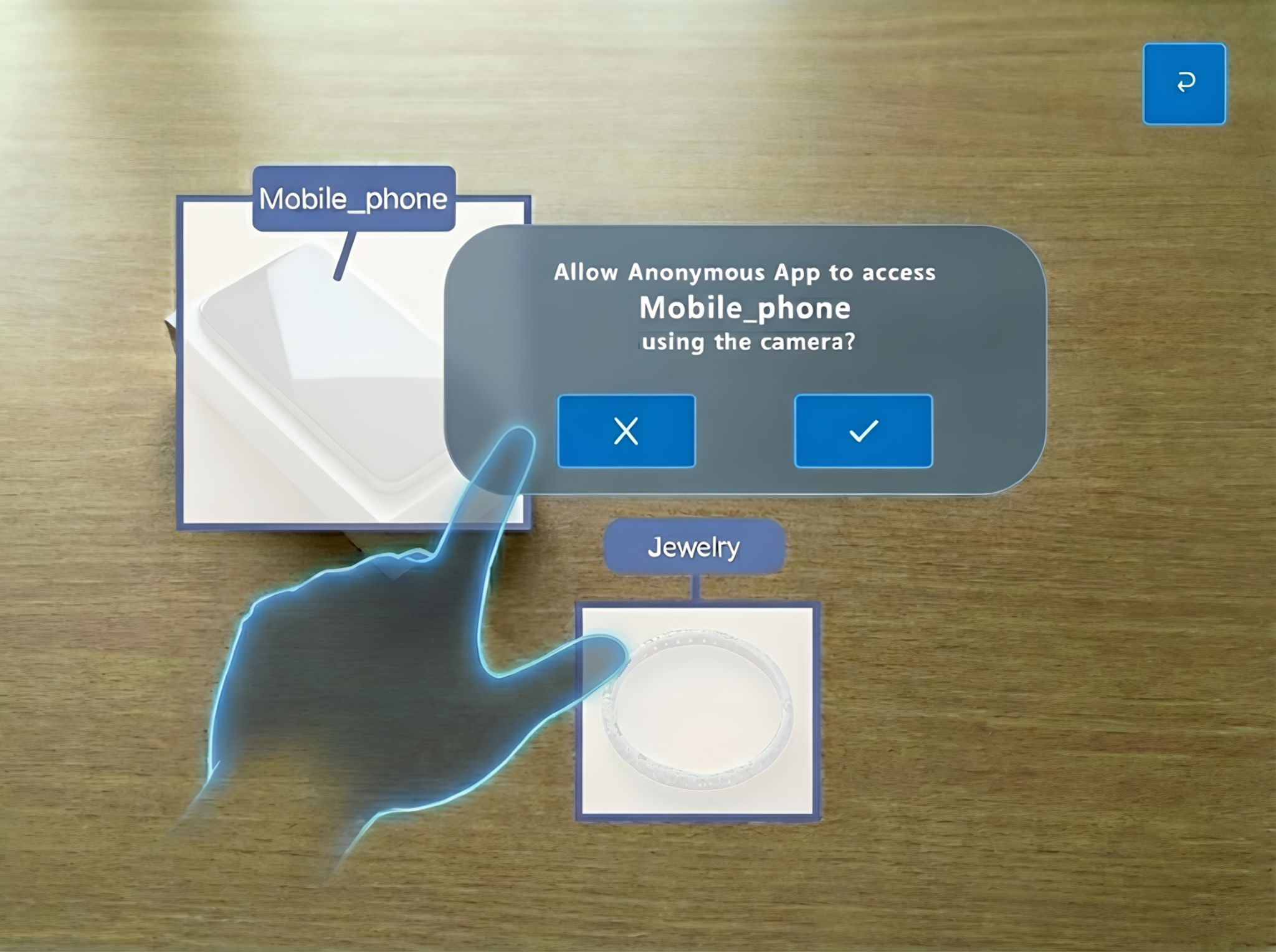}
    }
    \subfloat[Slider-based.]{
        \includegraphics[height=0.22\textwidth]{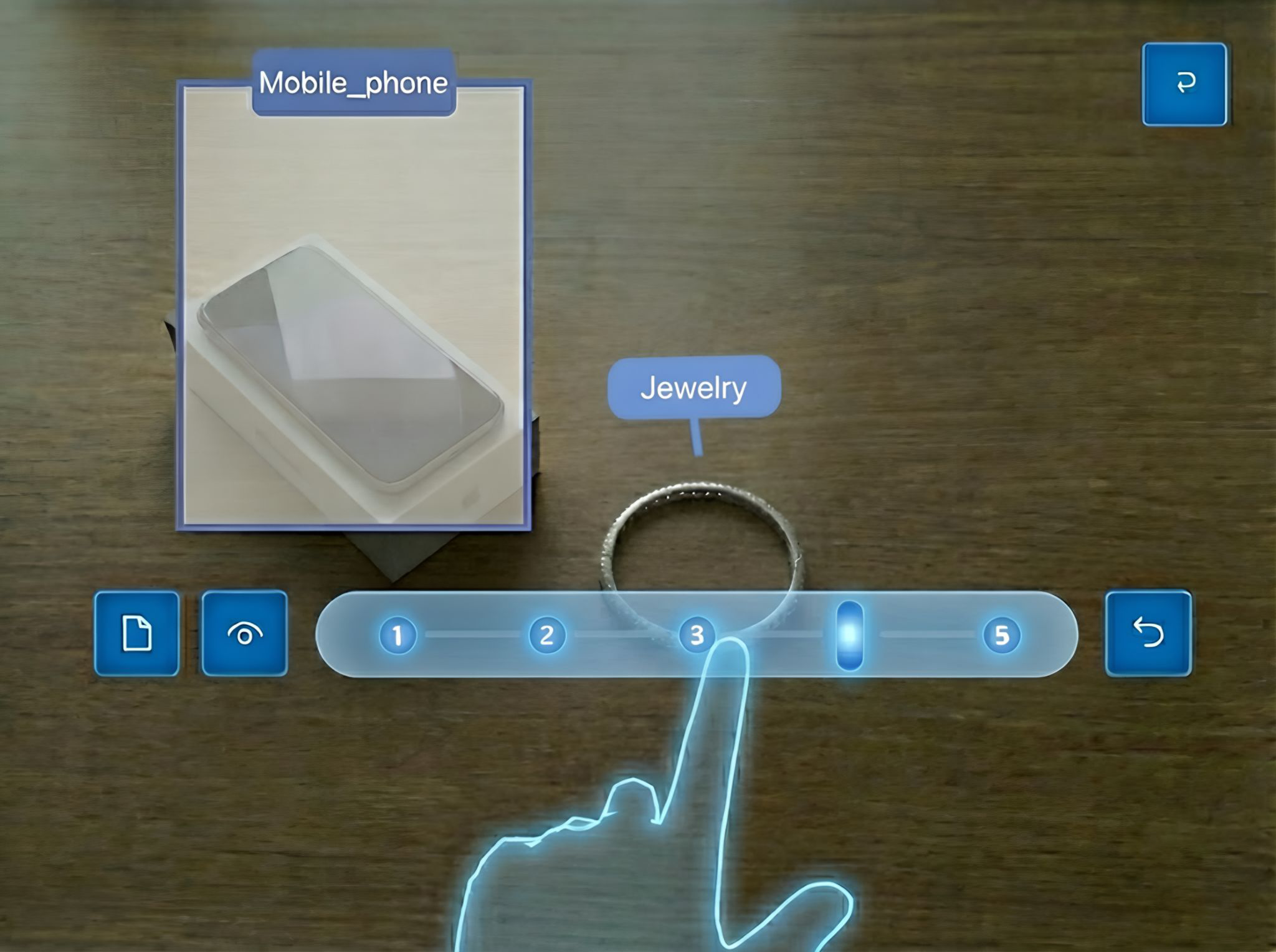}
    }
    \subfloat[VisGuardian.]{
        \includegraphics[height=0.22\textwidth]{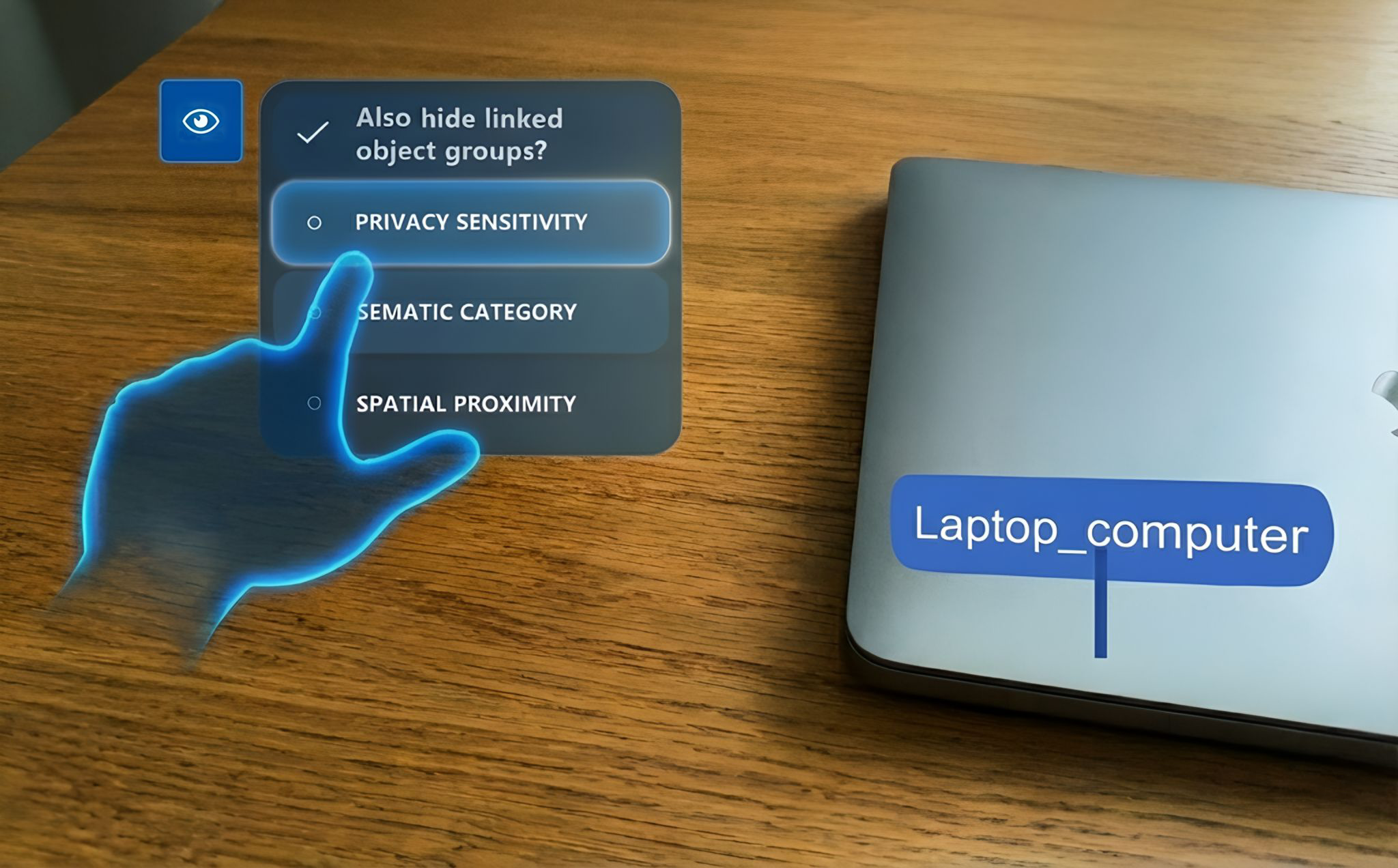}
    }
    \caption{The illustration for different techniques.}
    \label{fig:interface}
\end{figure*}

\begin{figure*}[!htbp]   
    \centering 
    \subfloat[Select task.]{
        \includegraphics[width=0.312\textwidth]{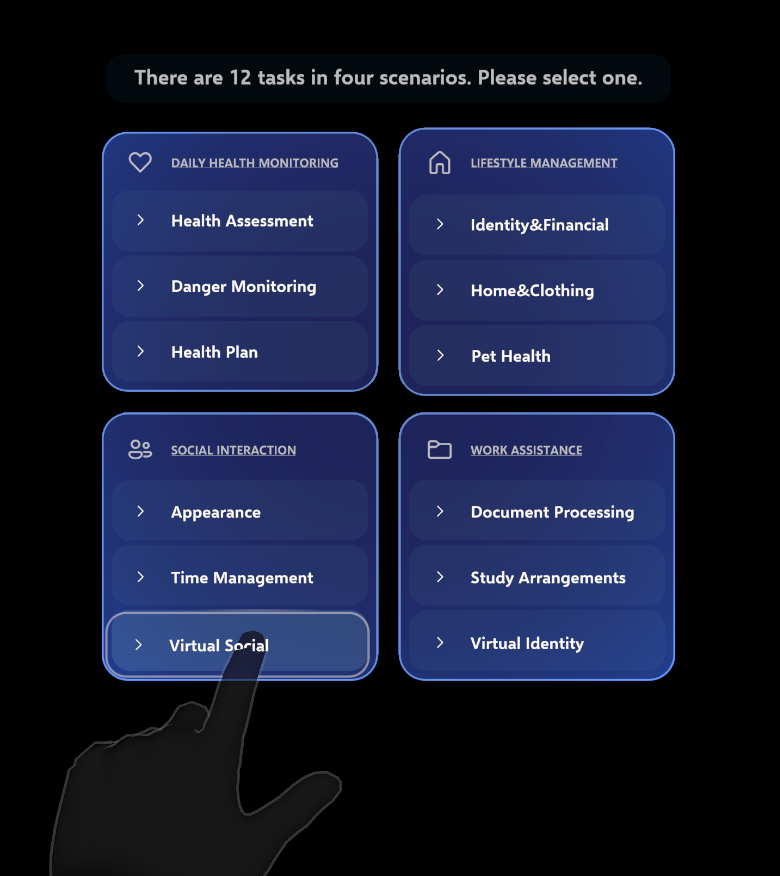}
    }
    \subfloat[Input voice and then click.]{
        \includegraphics[width=0.34\textwidth]{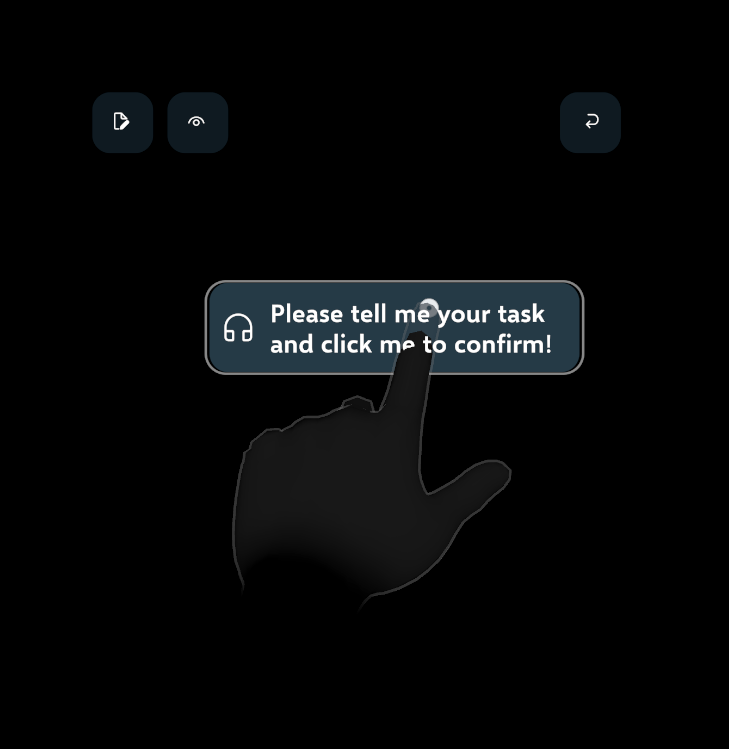}
    }
    \subfloat[View answers.]{
        \includegraphics[width=0.285\textwidth]{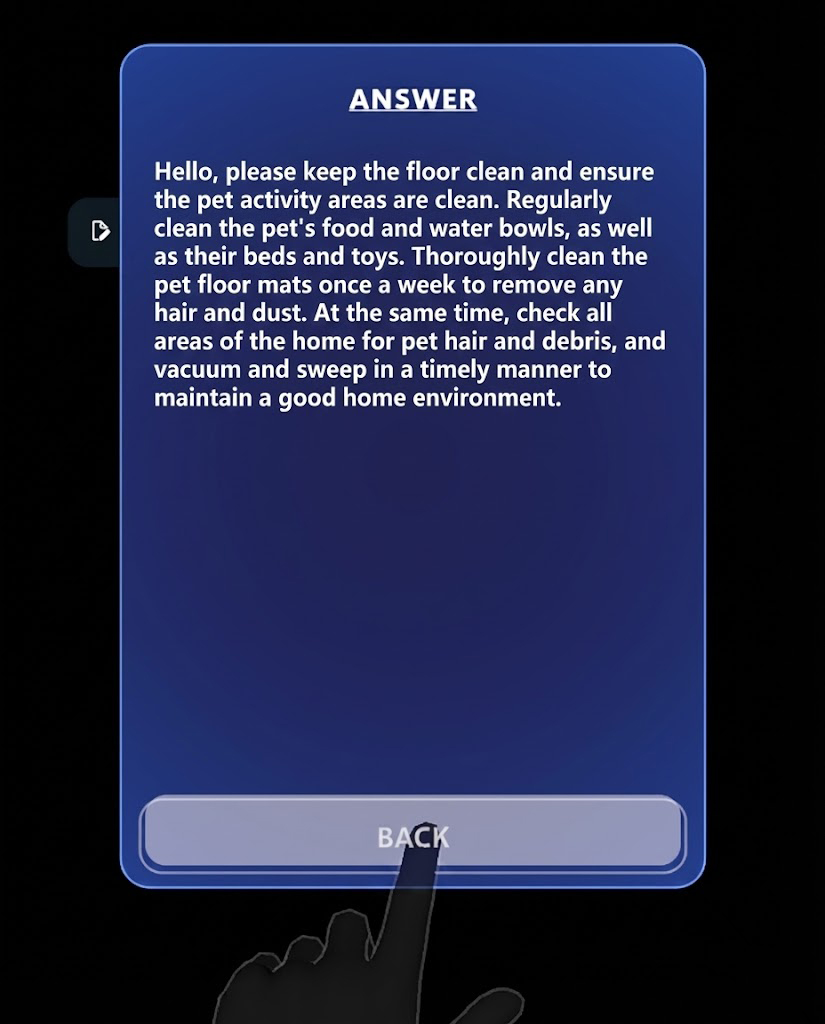}
    }
    \caption{The illustration of the interfaces for the experiment, where (a) the user select the task from the corresponding scenario, (b) the user input voice and then click to proceed, (c) the user viewed the answer.}
    \label{fig:interface_illustration}
\end{figure*}

We selected 12 specific tasks divided across four representative domestic scenarios (see Table~\ref{tab:task_scenarios}). These scenarios were grounded in prior literature on domestic privacy and AR utility~\cite{arakawa2023prism,arakawa2024prism,steil2019privaceye,mahmud2024actsonic}. The tasks were iteratively refined by six researchers to ensuring they realistically necessitated AR and AI applications' assistance while presenting distinct privacy risks. Participants performed tasks simulating daily routines within a home setting by reading the task prompt, immersing themselves in the scenario and using AR to execute the corresponding actions (see Figure~\ref{fig:interface_illustration}).

\begin{table}[!htbp]
    \centering
    \caption{Detailed breakdown of the 12 experimental tasks across four domestic scenarios. Each scenario consists of three distinct tasks designed to introduce specific privacy risks and AR interaction needs.}
    \label{tab:task_scenarios}
    \scriptsize 
    \begin{tabularx}{0.5\textwidth}{llX}
        \toprule
        \textbf{Scenario} & \textbf{Task} & \textbf{Task Description} \\
        \midrule
        
        \textbf{Daily Health \&} & 1. \textbf{Morning Check-up} & Analyzing facial features in a bathroom mirror to assess mental state and suggest care routines. \\
        \textbf{Behavior} & 2. \textbf{Dietary Monitor} & Recognizing food types and intake amounts during meals to offer nutritional advice.\\
        & 3. \textbf{Sedentary Alert} & Monitoring posture and screen focus during work to suggest physical breaks. \\
        \midrule
        
        \textbf{Household \&} & 4. \textbf{Cooking Assistant} & Identifying ingredients and tools to guide culinary steps and recipes. \\
        \textbf{Lifestyle} & 5. \textbf{Cleaning Guide} & Detecting dirty areas (e.g., stains) and recommending cleaning tools/steps. \\
        & 6. \textbf{Inventory Check} & Scanning refrigerator contents to track stock and identify expiring items. \\
        \midrule
        
        \textbf{Social} & 7. \textbf{Guest Insight} &  Analyzing a guest's facial expressions during a house party for advise on social etiquette. \\
        \textbf{Interaction} & 8. \textbf{Conversation Aid} & Monitoring tone and engagement in a 1-on-1 private chat to improve communication. \\
        & 9. \textbf{Interaction Summary} & Generating a post-event report on a social event happened in the home. \\
        \midrule
        
        \textbf{Work} & 10. \textbf{Schedule Sync} & Visualizing calendar events and to-do lists on the desk surface. \\
        \textbf{Assistance} & 11. \textbf{Meeting Scribe} & Recording audio/video of a remote meeting to extract key minutes and action items. \\
        & 12. \textbf{Task Tracking} & Monitoring workflow progress and documents to provide productivity reminders. \\
        
        \bottomrule
    \end{tabularx}
\end{table}

To simulate a functional AI assistant for our user study, we integrated a Vision-Language Model as the interactive backend. Specifically, we utilized OpenAI's GPT-4o via its API. During each task, the sanitized video frame, along with their spoken query, was sent to GPT-4o upon the user's querying. The model's objective was to generate a contextually relevant, conversational response, which was then shown to the participant. The system prompt used to guide the model's behavior is provided \textit{in the supplementary materials}. This methodology allowed us to evaluate user interactions with a state-of-the-art AI system in a controlled setting. For evaluation, we measured the following aspects and conducted audio-recorded semi-structured interviews (scripts in Appendix~\ref{app:script}): 

\textit{\textbf{Permission Control Time (corresponding to H1, efficiency)}}: The time elapsed, in seconds, from the moment the permission-setting interface was displayed until the user confirmed their settings to begin the main task~\cite{habib2022evaluating}.

\textit{\textbf{Number of Clicks (corresponding to H1, efficiency)}:} The total number of operations recorded during the permission control time~\cite{habib2022evaluating}. This is an aggregation involving individual data objects, distinct pointing of positions on a slider, and interactions with any interactive element within each widget for each technique.

\textit{\textbf{Perceived privacy protection and task performance~\cite{zhang2024adanonymizer} (corresponding to H2, usability):}} we measured the perceived privacy protection effectiveness and task performance.

\textit{\textbf{Overall perception~\cite{abraham2024you} (corresponding to H2, usability):}} we measured the satisfaction of participants using the scale \textit{``The permission control method provided the opportunities to set the best privacy decision''.}

\textit{\textbf{Workload~\cite{abraham2024you} (corresponding to H2, usability):}} Inspired by NASA-TLX~\cite{hart1988development}, we used its six dimensions to measure users' perceived workload. Consistent with practices in HCI research~\cite{cai2021makes,zhang2023concepteva}, we converted these scales to 7-point Likert scales to align with other dimensions' measurements. 

\textit{\textbf{Ease of use~\cite{abraham2024you} (corresponding to H2, usability):}} we measured the ease of use by~\cite{brooke1996sus} \textit{``I thought the permission control method was easy to use''}.

\subsection{Procedure}

Participants provided informed consent before beginning the study. The initial training protocol introduced them to the experimental platform (HoloLens 2) and demonstrated the three anonymization techniques: VisGuardian, Object-based, and Slide-based. To mitigate initial learning effects and ensure familiarity, participants were given three minutes to freely practice the anonymization effects of each technique before the main study. The main study comprised three sessions, one for each technique, separated by brief breaks. Each session featured four distinct scenarios (one task per scenario). Both the sequence of techniques and the presentation order of scenarios within each session were counterbalanced using a Latin Square design to control for order effects. The total duration of the study, including training and the three main sessions, but excluding the 15-minute exit interview, was approximately 60 minutes.

In each session, participant wore AR glasses and received experimenter guidance. Their primary aim was to complete a task while concurrently controlling their visual privacy with the assigned technique. The task required participants to use the AR glasses to view specific objects as contextual input, vocalize the task, click a button to query, and receive the answer--mimicking typical AR glass usage~\cite{MetaOrion2024}. To adjust privacy settings, participants could directly point or touch to manipulate the objects. They were permitted to redo the task or modify the privacy settings if they were unsatisfied with the task result. After each session, participants completed a questionnaire and have an additional 90-second break. After the three sessions, they participated in an exit interview. 

\subsection{Data Analysis}

We used Repeated Measures Analysis of Variance (RM-ANOVA) for the analysis of quantitative data and Friedman non-parametric tests for the analysis of subjective ratings. We performed Tukey Honestly Significant Difference (Tukey-HSD) test and Nemenyi test (with Bonferroni adjustment) for post-hoc comparisons. All significances were reported at $p < .05$. For interviews, one author transcribed the recordings and the other author checked for correctness. We adopted thematic analysis~\cite{braun2006using} on transcribed scripts, adopting an inductive coding strategy to identify emerging patterns without pre-conceived categories. Two authors first separately coded 4 scripts to establish the initial codebook. They discussed to resolve disagreements, with an inter-rater reliability of 0.81 calculating Cohen's Kappa. They then separately coded half of the remaining scripts, iterating the codebook and discussing intermittently if there were disagreements. After coding all the corpus, they performed axial coding to aggregate codes into themes based on semantic similarity. Frequencies were calculated post-coding. Quotes presented in the findings are verbatim, with minor grammatical corrections for clarity. The final codebook is detailed in Appendix~\ref{app:codebook}.

\subsection{Results}

\textbf{For battery consumption and latency,} we observed that during the 60-minutes' experiment, participants consumed an average of 35.0\% battery with \proj{} while consuming 35.2\% and 35.5\% battery for slider-based and object-based techniques. Experimenters also tested that without loading the model, the 60 minutes' experiment consumed 33.3\% battery. Different techniques did not have significant differences in battery consumption. We logged the latency and found on average, the utilized Yolov10n model resulted in 14.0 ms on-device latency on the HoloLens 2, including both forward pass and post-processing, which is feasible for on-device deployments.

\subsubsection{Efficiency: Permission Control Time and Clicking Times}

\begin{figure}
    \includegraphics[width=0.5\textwidth]{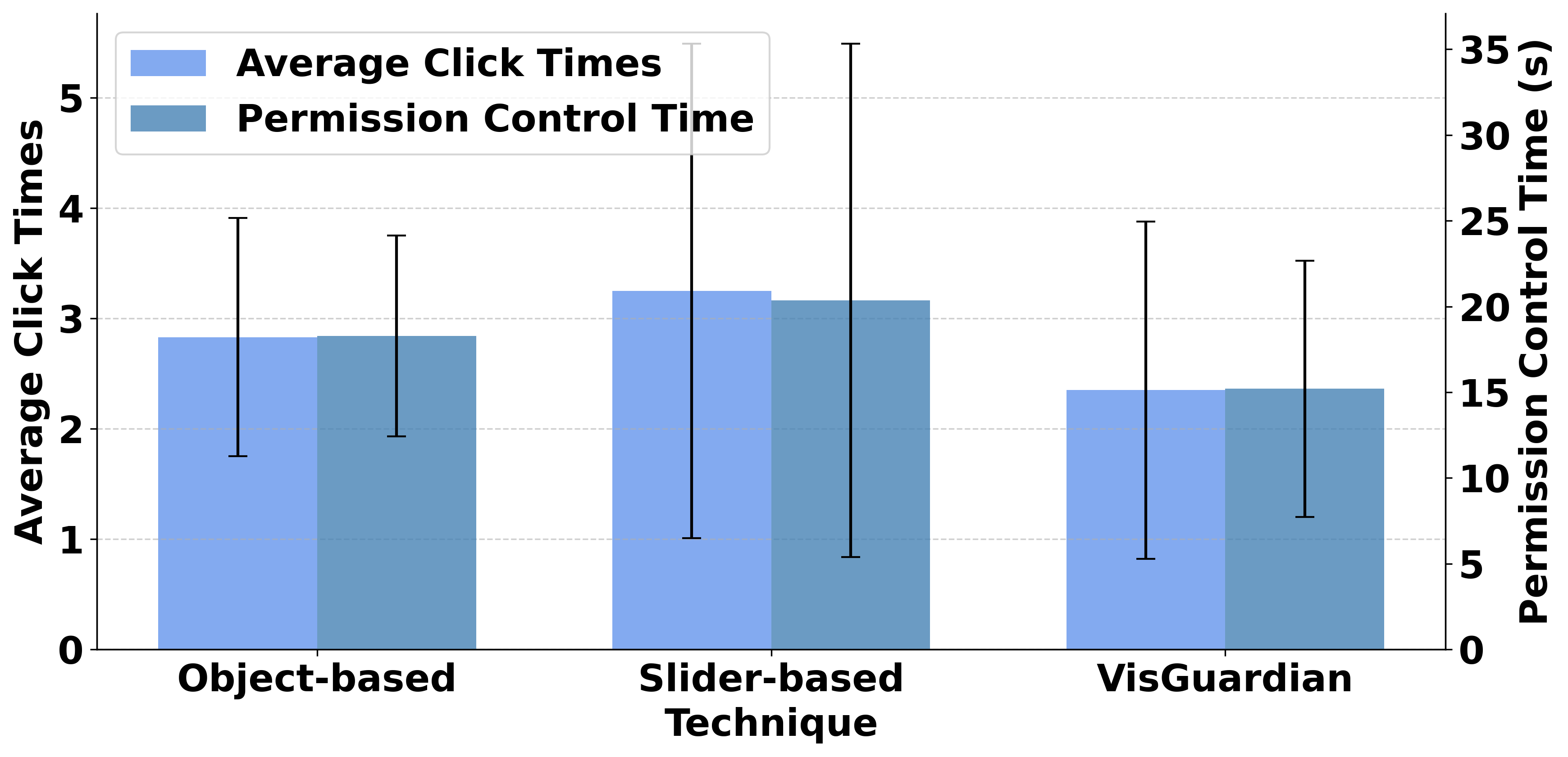}
    \caption{The average number of clicks and permission control time of different techniques. Errorbar indicated one standard deviation.}
    \label{fig:time}
\end{figure}

Figure~\ref{fig:time} shows the permission control time and clicking times for each scenario and task. We observed that \proj{} outperformed other techniques in terms of permission control time and clicking times. These results demonstrate the efficiency of \proj{}. Specifically, the average permission control time of \textit{\proj{}}, \textit{slider-based control} and \textit{object-based control} is 15.20s (SD=7.47s), 20.36s (SD=14.96s) and 18.28s (SD=5.85s) respectively ($F_{2, 46} = 4.034$, $p < .05$, post-hoc comparisons $p < .05$ between \proj{} and \textit{slider-based control}). We also observed a significant main effect of the technique on clicking times ($F_{2,46}=4.247$, $p<.05$), indicating an overall difference in user interactions across the three conditions, though post-hoc pairwise comparisons did not find a significant difference between any two techniques. 

Notably, \proj{} achieved a shorter total time per task (M=120.6s, SD=105.3s) compared to \textit{slider-based control} (M=141.5s, SD=110.8s) and \textit{object-based control} (M=135.9s, SD=87.3s) ($F_{2,46} = 4.547$, $p < .05$), which proved its overall effectiveness. Furthermore, we did not find significant effect of scenarios on users' permission control time for different techniques ($p = .75$, $p = .28$, $p = .44$ respectively), probably due to the fact that the distribution of privacy objects was rather balanced in different scenarios.

These quantitative findings are substantiated by our qualitative analysis, which shows \textit{how} the group-based mechanism (e.g., groupings, ungroup operations, single-object operations) lowers interaction cost despite the introduction of a selection menu: 

\textbf{1. Reduction of Repetitive Operations:} Users reported that the group-based control reduced operations in cluttered environment. Instead of performing discrete operations for each sensitive objects, participants utilized \proj{} to sanitize entire categories via a single interaction sequence. P12 highlighted this efficiency, noting that \proj{} helped \textit{``automatically erase similar and sensitive objects faster than other techniques''} 

\textbf{2. Minimization of Cognitive Load:} Compared to the \textit{slider-based} setting, which users described as an \textit{``opaque mechanism''} requiring trial-and-error to determine which objects were hidden (P5), \proj{} offers transparent control. The group selection provides a clear semantic mapping, eliminating the cognitive burden of exploring slider levels.


\subsubsection{Permission Setting Strategy}\label{sec:strategy}

Through interviews, we identified three dominant operational strategies for \proj{}'s permission control: 

\textbf{Select upon request:} 13/24 participants configured the permission before completing the task, upon initial settings. By leveraging the group-based operations, users configure the environment in a single pass (e.g., selecting one object and then configure groups). This was particularly effective in static home environments, where a single group setting persisted across the session, obviating the need for continuous manual adjustments common in other alternative settings.

\textbf{Preview when needed:} 12/24 participants praised the feature to click and preview the privacy protection during the task execution, although they may not change the settings. To them, checking the privacy status is \textit{``as easy as checking the remaining battery.'' (P2)} \textit{``Privacy to me is similar to something like battery, I do not want battery to drain out, I also do not want privacy to go unprotected. Thus I would intermittently spend a second to check the privacy protection status, especially when there is new objects emerging.'' (P16)} They argued that owing to the clear visualization strategy, they could quickly get a sense of which group(s) of information are anonymized.

\textbf{Hybrid Control with Granular Refinement :} 15/24 participants mentioned a hybrid strategy, utilizing group selection for bulk processing, while sometimes reserving manual selection for specific exceptions. While the group mechanism memorized and handled the majority of sensitive items (P14, P15), saving users' time by avoiding object-by-object selection, \proj{} supported necessary adjustments. In a few complex scenarios, users used additional steps for granular tuning. For example, as P16 explained, \textit{``On a few occasions, I have different objects that needed to be hidden, and then I need two or more operations.''}

Regarding the specific permission control settings and choices, we identified three patterns:

\textbf{Utility-Driven Permission Granting:} 23/24 participants chose to reveal sensitive objects when they were functionally required for task completion, performing a privacy calculus~\cite{laufer1977privacy}. In this strategy, the immediate utility of the AI assistant superseded privacy concerns. For instance, participants voluntarily disabled the default obfuscation for high-risk objects like screens during work-related tasks to enable text recognition. P11 articulated this trade-off: \textit{`` ... for this scheduling task, I needed to see the details ... You have to let it see some things ...''} This dynamic highlights the limitations of static permission models and the importance of fine-grained controls like \proj{}. 

\textbf{Personal and Social Attribute Management:} 15/24 participants maintained obfuscation settings not just to protect Personal Identifiable Information (PII), but to actively avoid leaking evidence of private habits. Their choices to obscure persons (e.g., themselves) were correlated with the reluctance to leak negative personal attributes (e.g., disorder or laziness) or intimate lifestyle details. As P18 marked, \textit{``hiding things like the stack of unread email [on the screen] ... It just feels too personal ...''} This indicates permission choices are employed to sanitize personal and social context rather than simply PIIs.

\textbf{Obfuscation due to uncertainty about third-party access:} 14/24 participants adopted a defensive strategy, maintaining the obfuscation states of objects that they thought marginally sensitive, due to the uncertainty regarding data flows. Lacking clear knowledge of downstream data usage, these users default to over-sanitization for privacy protection. P6 expressed this rationale: \textit{`` ... who knows where this video ends up? I'd rather erase more things ... just to be private.''} This highlights users' permission choices tend towards mitigating speculations from unknown data practices.

\subsubsection{Subjective Ratings and Suggestions for Improvements}\label{sec:subjective}

\begin{figure*}
    \includegraphics[width=0.9\textwidth]{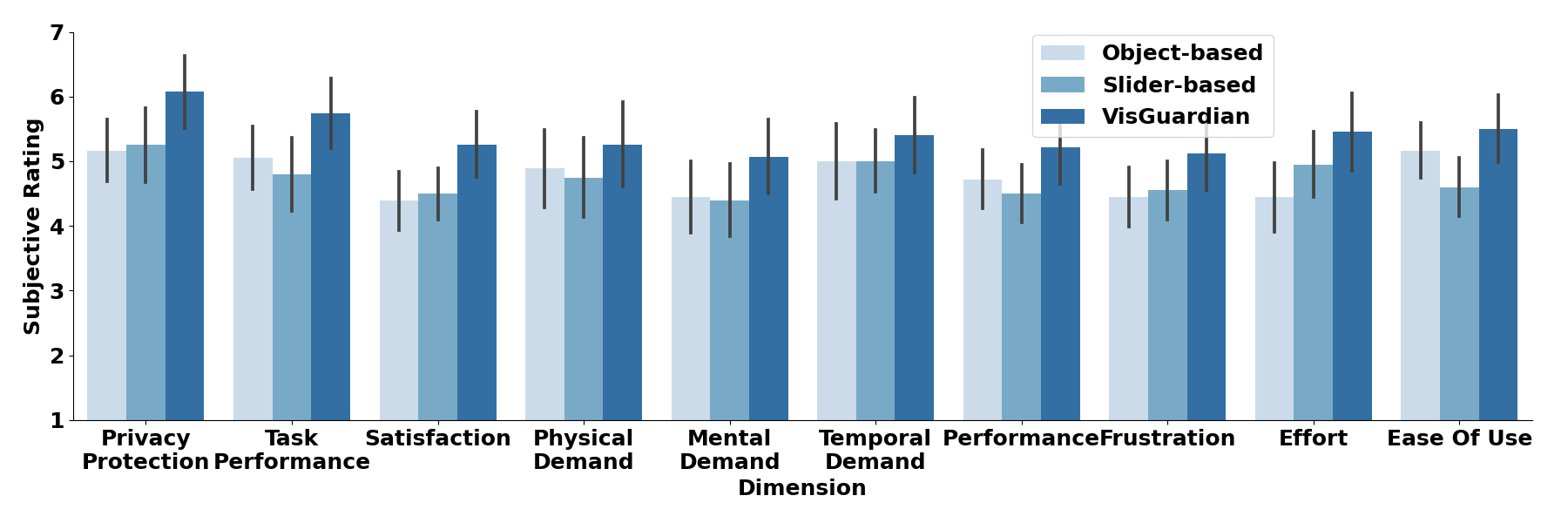}
    \caption{Barplot showing the subject ratings of participants (1: most negative, 7: most positive). Physical/mental/temporal demands, frustration, and effort scores were reversed (just in the paper for presentation) so higher values universally indicate more positive outcomes (e.g., lower demand). Errorbar indicated one standard deviation.}
    \label{fig:subj}
\end{figure*}

Figure~\ref{fig:subj} showed participants' subjective feedback, where \proj{} outperformed other techniques in terms of privacy protection, effectiveness, performance and ease of use (Privacy protection: $\chi^2_2 = 7.131$, $p < .05$; Effectiveness: $\chi^2_2 = 10.008$, $p < .01$; Satisfaction: $\chi^2_2 = 3.100$, $p = .21$; Physical effort: $\chi^2_2 = 3.265$, $p = .20$; Mental effort: $\chi^2_2 = 3.196$, $p = .20$; Time cost: $\chi^2_2 = 3.299$, $p = .19$; Performance: $\chi^2_2 = 6.950$, $p < .05$; Frustration: $\chi^2_2 = 3.762$, $p = .15$; Difficulty: $\chi^2_2 = 0.620$, $p = .73$; Ease of use: $\chi^2_2 = 9.287$, $p < .01$; all post-hoc $p < .05$ compared to other techniques.). While for other dimensions, \proj{} did not show significant advantages over other techniques. Participants commented that \proj{} was easy to understand even for novices to AR glasses, and \textit{``I do not need to spend time getting acquainted with that. Visualizing and clicking are so natural that it is quite like on the screen.'' (P10)}. \textit{``I think \proj{} is intuitive and the concept of grouping is natural and convenient.'' (P15)}

Participants also suggested areas of improvement, which we summarized below: \textbf{1. Customized Privacy Objects} A few participants mentioned the need for defining privacy objects by themselves, such as \textit{``the text on the magazine of documents'' (P1)} or \textit{``objects which could indicate I am not married and lived single.'' (P10)} This shows a demand for granular control over items whose privacy relevance stems not only from inherent PII but also from their potential to imply sensitive personal states or attributes.

\textbf{2. Articulation of the Task's Need} The current version of \proj{} did not consider the specific permission needs of the task. Some users require explicit rationale to make informed privacy trade-offs. As articulated by P8, \textit{``I want to know why the apps wanted to collect these objects, so I could decide better, rather than just based on my self-assessment.''}

\section{Discussions and Design Implications}

Our findings demonstrate the practical feasibility of \proj{} as an on-device privacy protection technique and highlight its user-centric interaction paradigm. This analysis leads to key design implications for future privacy-preserving techniques.

\subsection{Feasibility of \proj{}}

\proj{}'s feasibility is grounded in its efficient technical implementation and its robust design for handling real-world imperfections. Its reliance on a close-set detection model meets the computational constraints of AR glasses, achieving low latency (14.0 ms) and a minimal model size (2.3 M)~\cite{kress2014segmentation}. Unlike systems that focus on single salient objects~\cite{venugopalan2024aragorn}, \proj{}'s ability to detect multiple contextual objects simultaneously is crucial for modeling the complex privacy-utility trade-offs inherent in real-world scenarios~\cite{abraham2024you,zhang2024s}.

No detection algorithm is infallible; therefore, \proj{} is designed to manage such failures from a usability perspective. In cases of false negatives (i.e., a failure to detect a private object), we observed that users intuitively adjust their viewing angle to assist the system. While effective, this interaction imposes a physical cost. We acknowledge this limitation and identify algorithmic enhancement as future work. Conversely, the cost of correcting false positives (i.e., a benign object being misidentified as private) is minimal, requiring only a single tap to reveal the object.

This interaction design ensures \proj{} is a viable system-level service. It can be integrated into platforms like UWP for AR glasses or adapted as a middleware layer for other camera-equipped devices like AI glasses, as long as they feature visual displays as feedbacks~\cite{venugopalan2024aragorn}. Its group-based selection principles could also be extended to screenless AI glasses and other wearables, albeit that may need alternative feedback modalities, such as haptic or audio cues~\cite{song2020m,mehta2021privacy}. This design offered a standardized approach to privacy management.

\subsection{Effectiveness of Group-based Mechanism}

The core of \proj{} lies in its group-based mechanism, effectively merging the scalability of class-based privacy rules~\cite{kim2023erebus} with the contextual flexibility of object-based interaction~\cite{venugopalan2024aragorn}. While accessing a menu introduces a minimal step for a single instance, this mechanisms substantially reduces the cumulative effort in realistic, object-dense home environments. By selecting a single anchor object, users can simultaneously apply privacy settings to all semantically or spatially related items (e.g., ``hide all documents'' or ``mask bedroom area''), thereby replacing repetitive individual selections with a single collective action. Our empirical results (Section~\ref{sec:user_evaluation}) support this design, showing that \proj{} reduced both task completion time and click counts compared to object-based settings.

Furthermore, our design prioritizes binary visibility control over granular abstraction (e.g., sliders~\cite{abraham2024you}). Although privacy is often conceptualized as a spectrum~\cite{nair2023going,abraham2024you}, prior research indicates that mapping abstract levels into privacy controls imposes difficulty in understanding~\cite{colnago2022concern}. Participants in our study also confirmed that slider-based controls requires cumbersome exploration. By replacing abstractions with group-based familiar controls, \proj{} reduces operational burden, allowing users to navigate the privacy-utility trade-off.

\subsection{Proactiveness in Permission Settings}

Permission systems for AR glasses can vary in proactivity from fully reactive to fully automated~\cite{zhang2019proactive,zhang2025situguard}. In reactive systems, users manually adjust privacy settings when a potential risk is detected~\cite{venugopalan2024aragorn}. This offers full control but is burdensome~\cite{bugiel2013flexible}, requiring constant attention and leading to user fatigue, especially in dynamic, real-time contexts like those with AR glasses. Fully automated systems predict and address privacy risks autonomously~\cite{yang2024feasibility,zhang2025evaluating,zhang2025towards}, minimizing cognitive load and interruptions. However, they may misidentify risks or lack user control, leaving users feeling disconnected from the system's decisions.

\proj{} balances these extremes with a proactive approach that automatically detects privacy risks but allows user refinement when necessary, reducing effort while maintaining control over complex scenarios, aligning with ``privacy-by-design''~\cite{cavoukian2009privacy}. An automated version of \proj{} could predict privacy needs based on ML algorithms, offering seamless protection but risking over-cautiousness or undetected risks. A reactive model would give more control but introduce high cognitive load, potentially disrupting tasks. Further automation or reactivity would need to consider user preferences, address emerging privacy risks ,and maintain user trust, which we leave for future work.

\subsection{Generalizability}\label{sec:generalize}

We evaluated \proj{} in home environments, a critical scenario for AR glasses~\cite{cai2025aiget}. However, the core concept of group-based control holds potential for broad contexts, such as offices or public spaces~\cite{aiordachioae2019life}. Generalizing to these settings would require fine-tuning detection models to recognize context-specific sensitive objects. The relatively low density of private objects in outdoor or self-logging scenarios also requires future work to validate group-based control's performance. 

Regarding device applicability, while \proj{} targets AR glasses, the proposed interaction paradigms may inform privacy controls for VR headsets~\cite{abraham2024you}, smartphones~\cite{venugopalan2024aragorn} or lifelogging devices~\cite{price2017logging} that need continuous visual sensing. For always-on sensing scenarios where traditional permissions struggle~\cite{o2023privacy}, our group-based approach may offer a viable framework. Extending this to screenless devices would necessitate replacing visual overlays with alternative feedback modalities, such as auditory~\cite{song2020m} or haptic cues~\cite{mehta2021privacy}, to maintain user awareness and control.

Finally, we acknowledge that visual privacy is culturally dependent~\cite{li2017cross,xu2024dipa2}. While prior research indicates consistency in high-risk domestic objects across cultures~\cite{wang2023modeling,wu2020privacy}, specific sensitivity levels and acceptability thresholds vary globally~\cite{abdi2021privacy,xu2024dipa2}. Consequently, cross-cultural deployment would require re-calibrating the privacy taxonomy and default groupings to align with local norms and regulations.

\subsection{Design Implications}

Our design of \proj{} offers several design implications regarding the private object visualization, group-based control, and privacy-utility balancing.




\textbf{Implication 1: Frame-based overlaid visualization as the basis for control in always-on sensing environment. } \proj{}'s frame-based overlay visualization proved effective for conveying privacy status and enabling direct manipulation of detected objects. This approach, which visually highlights detected sensitive objects directly on the camera feed, can be adapted for other camera-based ubiquitous computing applications (e.g., lifelogging devices~\cite{price2017logging}). It provides users with immediate, contextual awareness of what the system sees and allows for intuitive, object-level control over privacy settings.

\textbf{Implication 2: Group-based control to ease user operations and cognitive load.} The study demonstrated that \proj{}'s group-based control mechanism significantly reduced configuration time compared to object-selection and slider-based methods. Grouping objects by privacy sensitivity, object category, or spatial proximity aligns with users' mental models and streamlines the process of applying privacy settings to multiple objects simultaneously. This metaphor offers a promising direction for managing fine-grained permissions efficiently in other systems where users might otherwise face high cognitive or operational load, such as complex mobile~\cite{venugopalan2024aragorn,price2017logging} or AR permission interfaces.

\textbf{Implication 3: Content-based permission control to facilitate users' privacy-utility balancing.} The effectiveness of \proj{} underscores the need to move beyond coarse, sensor-level permissions towards content-aware controls. Designers should empower users with mechanisms to manage privacy with an object-group level, allowing for nuanced, context-dependent decisions about what visual information is shared or obscured. This approach supports dynamic privacy-utility balance, enabling users to permit necessary data for application functionality while protecting specific sensitive elements within the visual feed, which is critical for user acceptance and trust in always-on sensing systems~\cite{franzen2024communicating}.

\section{Limitations and Future Work}

We acknowledge several limitations of this work that suggest future directions. First, regarding study design and ecological validity, our participant sample was restricted and excluded demographics such as children and the elderly. While these groups may not be the primary operators of AR glasses, their presence in home environments warrants investigation into their interactions with \proj{}~\cite{bhardwaj2024focus}. Furthermore, the study employed a controlled environment with pre-placed objects, a common practice in prior work~\cite{venugopalan2024aragorn,abraham2024you}. Future work should conduct evaluations under varying environmental conditions. 

Second, \proj{} relies on a pre-defined, static taxonomy for privacy objects and mappings (e.g., associating nudity with persons), which currently precludes user-driven augmentation. As privacy definitions are often subjective and culturally contingent~\cite{xu2024dipa2}, these fixed mappings may not capture the full nuance of user preferences. Additionally, we adopted a static classification schema rather than dynamically grouping permissions and risks due to computational constraints. Implementing real-time, dynamic inference of contexts needs additional models and algorithms, which impose additional latency and power consumption overheads for edge devices. Future research could explore dynamic contextual reasoning to capture situational nuances such as distinguishing intentional user interaction and incidental background capture, thereby enabling more granular risk assessment.

Third, \proj{} focuses on the privacy of visual data captured by the front camera, resolving a significant risk in AR usage. Extending our group-based approach, future work could validate similar mechanisms on other pervasive sensors like microphones and depth cameras.

\section{Conclusion}
This paper addresses the privacy challenges posed by always-on visual sensing in AR glasses, where traditional permission models fall short. We introduced \proj{}, a system-level, content-based permission technique enabling fine-grained privacy control at the object-group level. \proj{} utilizes real-time object detection, classifying visual elements based on a synthesized privacy taxonomy, and offers intuitive, adaptive control via clicking-triggered, group-based selections (categorized by privacy sensitivity, object category, and spatial proximity). System evaluations demonstrated \proj{}'s efficiency and accuracy (0.6704 precision, 14.0 ms latency, 1.7\% battery impact). A user study (N=24) comparing \proj{} with slider-based and object-selection baselines across realistic scenarios confirmed its superiority. \proj{} significantly reduced configuration time while enhancing perceived privacy protection effectiveness, usability, transparency, and user satisfaction, effectively facilitating the balance between privacy preservation and task utility in AI-based AR applications.

\begin{acks}
    This work is supported by the Natural Science Foundation of China (NSFC) under Grant No. 62132010. 
\end{acks}

\bibliographystyle{ACM-Reference-Format}
\bibliography{sample-base}

\appendix 

\section{Ethical Considerations}

We acknowledged the potential ethical concerns and meticulously designed the experiments to address these concerns. We adhered to the Menlo and Belmont report~\cite{bailey2012menlo,beauchamp2008belmont} in designing the studies and our studies got the approval of our university's institutional review board (IRB). The private objects in the smart home environments were not participants' own objects but alternative objects for experimental use. Participants have the right to delete their own data or quit the experiment at any time during the experiment. We properly compensated participants according to the local wage standard. Our study aimed for developing an effective, user-friendly privacy control permission system, which benefited both research and academic fields. 

\section{Dataset Details}\label{sec:distribution}

According to the Table~\ref{tab:merged_privacy_objects}, and the distribution of the datasets' labels, we selected the following labels in the dataset with the following correspondence. We selected the classes for training as shown in Table~\ref{tbl:class_training}.

\begin{table}[!htbp]
    \centering
    \caption{The classes for training we selected.}
    \label{tbl:class_training}
    \begin{tabularx}{\textwidth}{llXl}
        \toprule
       Dataset Source & Original Class ID & Classes in LVIS/COCO & Class in our training \\ \midrule 
       LVIS & 792 & person/baby/child/boy/girl/man/woman/human & person \\ \hline 
       LVIS & 229 & cellular telephone/cellular phone/cellphone/mobile phone/smart phone & cell phone \\ \hline 
       LVIS & 126 & book & book \\ \hline
       LVIS & 884 & ring & ring \\ \hline
       LVIS & 145 & bracelet & bracelet \\ \hline
       LVIS & 147 & brassiere/bra/bandeau & bra \\ \hline
       LVIS & 526 & halter top & halter top \\ \hline
       LVIS & 753 & pantyhose & pantyhose \\ \hline
       LVIS & 748 & pajamas/pyjamas & pajamas \\ \hline 
       LVIS & 697 & monitor/monitor computer equipment & monitor \\ \hline  
       LVIS & 630 & laptop computer/notebook computer & laptop \\ \hline 
       LVIS & 111 & badge & badge \\ \hline 
       LVIS & 210 & identity card & identity card \\ \hline 
       LVIS & 1096 & toilet & toilet \\ \hline 
       LVIS & 1135 & urinal & urinal \\ \hline 
       LVIS & 236 & checkbook/chequebook & checkbook \\ \hline 
       LVIS & 437 & file cabinet/filing cabinet & file cabinet \\ \hline 
       LVIS & 682 & medicine & medicine \\ \hline
       LVIS & 641 & license plate/numberplate & license plate \\ \hline 
       LVIS & 1133 & underwear/underclothes/underclothing/underpants & underwear \\ \hline 
       LVIS & 1044 & swimsuit/swimwear/bathing suit/swimming costume/bathing costume/swimming trunks/bathing trunks & swimsuit \\ \hline 
       COCO & 0 & person & person \\ \hline 
       COCO & 67 & cell phone & cell phone \\ \hline
       COCO & 73 & book & book \\ \hline 
       COCO & 62 & tv & monitor \\ \hline 
       COCO & 63 & laptop & laptop \\ \hline 
       COCO & 61 & toilet & toilet \\ \hline
    \end{tabularx}
\end{table}

\section{The Home Environment}

Figure~\ref{fig:environment} showed the home environment we used. We also included different private objects common in the home environment in different places. The tasks took place in the natural place inside the smart home. The ``daily health monitoring and behavior tracking'' task takes place in the bedroom. The ``household and lifestyle management'' task takes place around the living room and the bathroom. The ``social interaction and contextual awareness'' and the ``multimodal learning and work assistance'' task takes place around the living room and the studying room separately. For the bedroom, the placed privacy objects included \textit{person, cellphone, book, ring, bracelet, bra, pantyhose, pajamas, laptop, identity card, medicine, underwear, swimsuit}. For the living room, the placed privacy objects included \textit{person, cell phone, book, ring, bracelet, halter top, pajamas, monitor, laptop, badge, identity card, file cabinet, license plate}. For the bathroom, the placed privacy objects included \textit{ring, bracelet, bra, pantyhose, toilet, urinal, underwear, swimsuit}. For the studying room, the placed privacy objects included \textit{person, cell phone, book, monitor, laptop, badge, identity card, checkbook, file cabinet, medicine, license plate}.

\section{The Tasks Selected in the Evaluation Study}\label{app:task}

\subsection{Daily Health Monitoring and Behavior Management}
This scenario involves health management, daily habit monitoring, and the collection of health-related data via the AI applications on AR glasses. These tasks often rely on AI to analyze the user’s health data, behavioral patterns, and physiological information. Privacy protection remains a key concern, particularly when the AI applications on AR glasses require continuous monitoring of a user’s physical or lifestyle habits. Example tasks include: health reminders and management, dietary behavior monitoring, lifestyle habit tracking.

\textbf{Privacy Issues}: There is a need to address how to protect users’ health data, dietary habits, exercise patterns, and other private information, ensuring the security of data storage and processing.

\subsubsection{Morning Health Management}

After waking up, the user prepares to wash up or has already done so. The AI applications on AR glasses detect the user's facial features to assess their mental state and emotions, suggesting personal care products and providing morning wake-up exercise recommendations. When the user starts the exercise program, the AI assistant analyzes the sensor data to evaluate the user's exercise status and offers advice.

\textbf{Privacy Content}: User's facial features, body shape, hairstyle, clothing, range of motion; personal care items.

\textbf{Scene Setup}: The room contains a mirror, yoga mat, and table, which is divided into two areas: one with personal care products such as facial cleanser, toothpaste, toothbrush, etc., and the other with personal items like a smartphone, headphones, books, cosmetics, and skincare products.

\subsubsection{Dietary Behavior Monitoring}

When the user is eating breakfast (or lunch, dinner), the AI applications on AR glasses analyze the types and amounts of food being consumed and offer health-related dietary suggestions.

\textbf{Privacy Content}: Tableware (chopsticks, knife and fork, plates), food, user’s hand and wrist accessories (rings, bracelets), clothing, and tablecloth patterns.

\textbf{Scene Setup}: A dining table with food for breakfast (or lunch, dinner), utensils, drinks, and a tablecloth. The user’s hands are holding utensils with accessories.

\subsubsection{Sedentary Behavior Reminder}

The user sits at a desk working, and the AI applications on AR glasses detect through sensors that the user has been sitting for a prolonged period. The AI assistant reminds the user to stand and move at regular intervals or tracks eye movements to determine if the user is looking at a computer screen, suggesting a break if needed.

\textbf{Privacy Content}: Computer screen information, tablet screen information, smartphone, personal items (e.g., water cups with text or patterns, handwritten notebooks, keyboard, mouse, mouse pad, desk nameplate, plants, posters, decor items like figurines, aromatherapy, vases), desk lamp, instant food like coffee powder.

\textbf{Scene Setup}: A desk with the aforementioned items.

\subsubsection{Evening Sleep Suggestions}

At night, the AI applications on AR glasses offer personalized sleep recommendations based on the user’s daily activity data, including exercise levels, dietary intake, emotions, and more. The user activates a relaxation audio application. 

\textbf{Privacy Content}: Bed linens, blankets, bedside lamp, smartphone, headphones, documents, books, and photos in the room.

\textbf{Scene Setup}: A room with a bed, bedside table, and related furniture.

\subsection{Household and Lifestyle Management}

These tasks involve how AI applications on AR glasses can assist users in efficiently completing household chores, such as cleaning, cooking, and laundry. Through AI integration, the glasses provide real-time feedback and guidance to improve household efficiency. Example tasks include: cooking assistance, cleaning management, household inventory management.

\textbf{Privacy Issues}: The user’s domestic environment and activities could involve the private lives of household members, raising concerns about data leakage, location tracking, etc. Therefore, safeguarding household environment and activity data is critical.

\subsubsection{Cooking Preparation}

The user enters the kitchen to prepare a meal. The AI applications on AR glasses display the recipe and cooking steps through AR, offering real-time advice based on the user's ingredients and tools. The user controls the glasses by eye movement (e.g., by focusing on an ingredient or tool), which shows the next step in the recipe and provides necessary cooking tips or reminders.

\textbf{Privacy Content}: Ingredient types, ingredient quantities, cooking tools, user’s facial features, hand movements, tabletop arrangement, kitchen environment items.

\textbf{Scene Setup}: A kitchen with a cooking counter, stove, refrigerator, kitchen tools (e.g., pots, knives, cutting boards, dishes), ingredients (e.g., vegetables, meat, spices), and possibly a recipe, phone, or other devices on the counter. The kitchen is well-lit, and the user stands near the workstation.

\subsubsection{Cleaning Task Management}

After cooking, the user decides to clean the kitchen. The AI applications on AR glasses, using built-in sensors and visual recognition, help the user identify areas that need cleaning and recommend appropriate cleaning tools. The AI assistant uses eye movement recognition to highlight the areas to be cleaned (e.g., stove, countertop) and provides cleaning steps and tool recommendations via voice instructions.

\textbf{Privacy Content}: User’s facial features, cleaning areas, kitchen surfaces, cleaning tools, cleaning actions, and user’s clothing (e.g., apron, gloves).

\textbf{Scene Setup}: Kitchen areas to be cleaned (e.g., stove, countertop, sink), cleaning tools (e.g., dish soap, cleaning cloths, sponges, gloves, mop), kitchen cabinets, sink.

\subsubsection{Inventory Management}

During the cleaning process, the AI applications on AR glasses detect that the user is missing certain ingredients (e.g., salt or oil) and remind them to replenish specific items. Through inventory tracking, the glasses also help manage the user's kitchen supplies. The user can query the stock of certain items (e.g., ``How much salt do we have left?''), and the glasses provide real-time data on the inventory and suggest shopping for items.

\textbf{Privacy Content}: Kitchen item types, item quantities, ingredient types, user's voice commands, user's facial features, the arrangement of user's items.

\textbf{Scene Setup}: Kitchen inventory area (e.g., cabinets, refrigerator, pantry shelves), ingredient packaging, containers, with the user standing next to shelves or a refrigerator. The glasses can scan and recognize items.

\subsubsection{Expiration Reminder for Household Items}

After dinner, the user decides to check the ingredients in the refrigerator. The AI applications on the AR glasses use the camera to scan the fridge’s contents, identify ingredients that are about to expire, and notify the user in time. The AI assistant provides alerts and suggests how to use or discard items. 

\textbf{Privacy Content}: Ingredient types, food packaging labels (date, brand), user's facial features, user actions (e.g., opening the refrigerator, retrieving items).

\textbf{Scene Setup}: Kitchen with refrigerator containing ingredients (e.g., vegetables, meat, beverages, spices), the refrigerator door is open, and the user stands nearby checking the contents.

\subsection{Social Interaction and Contextual Awareness}

This type of scenario focuses on how AI applications on the AR glasses can enhance the user's social interactions within the home (e.g., receiving visitors) by analyzing emotions, providing social cues, and optimizing communication. Tasks in this scenario may involve emotion recognition, facial expression analysis, and speech analysis. Example tasks include: emotion recognition and feedback, social AI assistant, personalized suggestions—based on the user’s social interaction patterns and preferences, the AI applications on the AR glasses provide personalized social advice or feedback (e.g., encouraging the user to engage in conversation with someone, recommending appropriate social activities, etc.).

\textbf{Privacy Issues}: Tasks involving emotion recognition and social behavior analysis may involve sensitive data. It is essential to ensure that the user's privacy is fully protected, especially regarding the leakage of personal emotional and psychological states.

\subsubsection{Guest / New Acquaintance Social Interaction Analysis and Advice}

The user interacts with a with a new acquaintance (e.g., a friend's partner or a distant relative) at a house party.. The AI applications on AR glasses use facial expression recognition and eye-tracking technology to analyze the guest’s emotions and offer real-time interaction advice to help the user optimize their communication strategy. 

\textbf{Privacy Content}: Guest’s facial features, emotional changes, eye contact, interaction between the user and the guest.

\textbf{Scene Setup}: A social gathering or party where the user is either standing or sitting next to a guest. The AI applications on AR glasses analyze the guest’s facial expressions and emotions, providing feedback.

\subsubsection{Emotion State Tracking and Improvement Suggestions}

During a one-on-one conversation with a friend or colleague, the AI applications on AR glasses monitor both individuals’ emotions through facial expression and voice analysis, offering suggestions to help regulate the interaction and improve emotional states.

\textbf{Privacy Content}: Facial expressions, voice tone, eye contact, body language, conversation content.

\textbf{Scene Setup}: The user and a friend or colleague are sitting indoors, having a casual or formal conversation. The AI applications on AR glasses track emotional shifts and offer suggestions in real-time.

\subsubsection{Post-Social Interaction Emotional Feedback and Optimization Suggestions}

After interacting with multiple people, the AI applications on AR glasses provide a summary of emotional changes throughout the conversation, offering a report and suggesting improvements for future social interactions to help the user enhance their communication strategy.

\textbf{Privacy Content}: Facial expressions, eye contact, emotional fluctuations, conversation content, voice tone.

\textbf{Scene Setup}: After a dinner or social activity, the user interacts with friends or colleagues. The AI applications on AR glasses provide an emotional analysis and feedback.

\subsection{Multimodal Learning and Work Assistance}

AI applications on AR glasses can help users in the workplace by enhancing productivity, managing complex tasks, and providing workflow support through AI technology. These tasks typically combine speech recognition, eye-tracking, and task automation to assist users in real-time operations or auxiliary functions. Example tasks include: task reminders and automation, meeting recording and collaboration, productivity enhancement.

\textbf{Privacy Issues}: Sensitive workplace data (e.g., company secrets, meeting content) must be strictly protected to prevent leakage or misuse.

\subsubsection{Task Scheduling}

The user starts their workday, and the AI applications on AR glasses automatically sync with their calendar to remind them of upcoming meetings or tasks. The AI assistant prepares the user by offering timely reminders of necessary materials or information, either through voice prompts or screen displays.

\textbf{Privacy Content}: Calendar entries, meeting schedules, document content, email content, user’s task list, work schedule, application access history.

\textbf{Scene Setup}: The user sits at their desk, with the AI applications on AR glasses displaying their calendar and reminders. The desk contains a computer, notebook, and work materials in a quiet office environment.

\subsubsection{Meeting Recording and Collaboration}

During a meeting, the AI applications on AR glasses use speech recognition and eye-tracking to automatically record meeting content and extract key information. The AI assistant provides keyword summaries or real-time notes to help the user follow up on meeting progress. The user can issue voice commands (e.g., ``Record this topic''), and the assistant will automatically label meeting notes and update in real time. 

\textbf{Privacy Content}: Meeting discussion content, confidential topics, participant identities, meeting duration, specific discussion details (e.g., company strategies, project progress), meeting recording data.

\textbf{Scene Setup}: A meeting room with participants seated around a table, the AI applications on AR glasses identifying and recording meeting content. The screen displays real-time notes and key information.

\subsubsection{Task Tracking}

After the meeting, the AI applications on AR glasses help the user track the progress of tasks derived from the meeting and provide automated reminders. Based on the user’s progress, the AI assistant periodically checks the status of the tasks and offers updates. The user can request to view task progress, and the assistant provides a task list and reminds them of the next steps.

\textbf{Privacy Content}: User’s task list, task progress, work goals, task priorities, related document content, task schedule.

\textbf{Scene Setup}: The user sits at their desk, using the AI applications on AR glasses to view their task list. The glasses display task status, deadlines, and priority information. The desk contains task materials, electronic devices, and work documents.

\subsubsection{Work Break Reminder}

The AI applications on AR glasses analyze the user's work behavior (e.g., fatigue, distractions) and offer suggestions to enhance productivity (e.g., take a break, refocus). Based on the data, the AI assistant offers efficiency-boosting tips (e.g., ``Your concentration time is over; take a 5-minute break'').

\textbf{Privacy Content}: User’s work behavior data, attention levels, fatigue levels, work duration, work frequency, and AI assistant’s optimization suggestions.

\textbf{Scene Setup}: The user sits at their desk, with the AI applications on AR glasses analyzing their work state and offering productivity suggestions. The desk contains a computer, work materials, and other office tools.

\section{The Mapping Relationship For the Slider-based Control In the User Evaluation}\label{app:mapping}

Table~\ref{tab:slider_mapping} outlines the mapping between slider positions and the specific object category selected for obfuscation. This hierarchy is grounded in the privacy risk classification detailed in Section~\ref{sec:categorization}. The control mechanism is cumulative: lower settings (left) mask only high-sensitivity attributes, while higher settings (right) progressively expand the scope to include objects of lower sensitivity. The scale ranges from full transparency (Position 1) to complete obfuscation (Position 5).

\begin{table}[htbp]
    \centering
    \caption{Mapping of objects to the control settings, where the object category denoted the object category to be obfuscated when the slider is at the corresponding position.}
    \label{tab:slider_mapping}

    \begin{tabularx}{\textwidth}{cX}
        \toprule
        Slider position & Object category to be obfuscated \\
        \midrule
        1 (obfuscate none) &  \\
        2 & person, medicine, underwear, license plate, jewelry \\
        3 & person, medicine, underwear, license plate, jewelry, toilet, mobile phone, laptop computer, gun, drunk \\
        4 & person, medicine, underwear, license plate, jewelry, toilet, mobile phone, laptop computer, gun, drunk, wheelchair, signed document, ID card, checkbook, swimsuit, calendar \\
        5 (obfuscate all) & person, medicine, underwear, license plate, jewelry, toilet, mobile phone, laptop computer, gun, drunk, wheelchair, signed document, ID card, checkbook, swimsuit, calendar, skirt, pajamas, legging, file cabinet, book, badge \\
        \bottomrule
    \end{tabularx}
\end{table}

\section{Interview Scripts For the Evaluation Study}\label{app:script}

We designed our questions to be open-ended and non-leading, while allowing participants to articulate their strategies and rationales without investigator bias. We encouraged them to ground their feelings and strategies in their experience. 

1. Could you describe your overall experience using this permission control technique?

2. What did you perceive as the primary advantages and disadvantages of using this method?

3. Can you walk us through the strategy you used to configure your privacy settings during the tasks?

4. How did you decide when to engage with the permission systems, such as at the beginning, during the task, after the task, or other occasions, or never?

5. What factors influenced your decision to hide or show specific objects in the environment?

6. Comparing different techniques, do you think they are the same of different? 

7. (If different) In what specific aspects, did you find them to differ?

8. Based on these comparisons, what techniques or what aspects of techniques aligns best with your personal preferences?

\section{Codebook For The Evaluation Study}\label{app:codebook}

Table~\ref{tab:qual_codebook} showed the codebook for the evaluation study.

\begin{table}[ht]
  \centering
  \caption{The final codebook applied in the thematic analysis. }
  \label{tab:qual_codebook}
  \small
  \begin{tabularx}{\textwidth}{@{}l l X@{}}
    \toprule
    \textbf{Theme} & \textbf{Code} & \textbf{Description \& Definition} \\
    \midrule

    \multirow{2}{*}{\textbf{Efficiency of VisGuardian}} 
    & Reduction of Repetitive Operation
    & VisGuardian reduces object-by-object operation times through group-based control. \\
    \cmidrule(l){2-3}
    & Minimization of Cognitive Load 
    & VisGuardian facilitates intuitive operations, which lowers users' load of understanding.\\
    \midrule
    
    \multirow{4}{*}{\shortstack[l]{\textbf{Operational Strategies}\\ \textbf{for Privacy Control}}} 
    & Proactive Group Configuration 
    & Users utilize group-based operations to sanitize the environment upon initialization. \\
    \cmidrule(l){2-3}
    & Intermittent Status Verification 
    & Users verify privacy visualization intermittently to check privacy state.\\
    \cmidrule(l){2-3}
    & Hybrid Refinement 
    & Users leverage group selection and then fine-grained selection to tune the settings. \\
    \midrule
    
    \multirow{3}{*}{\shortstack[l]{\textbf{Drivers of Permission}\\ \textbf{Decision Choices}}} 
    & Utility-Driven Granting 
    & Permission decisions are influenced by functional necessity, exhibiting privacy-utility trade-offs. \\
    \cmidrule(l){2-3}
    & Personal and Social Attribute Management 
    & Obfuscating decisions are motivated by personal impressions' management and social boundaries. \\
    \cmidrule(l){2-3}
    & Precautionary Defense 
    & Permission choices are influenced by a lack of trust or knowledge regarding downstream data usage. \\
    \bottomrule
  \end{tabularx}
\end{table}

\end{document}